\documentclass[%draft,
10pt, a4paper]{article}		%define general setup of document
\usepackage{titlesec}
\usepackage{grffile}
\usepackage{authblk}		% authors & affiliations
\usepackage[centering,hmargin=15mm,vmargin=2cm]{geometry}
\usepackage[labelfont=bf]{caption} % make figure caption in bold.
\usepackage[pdftex]{graphicx, xcolor}
\usepackage{siunitx}
\usepackage[backend=biber, natbib=true, style=nature, autocite=superscript]{biblatex}		%references (nature style)
\usepackage{amsmath,bm} % math mode
\usepackage{setspace}		% line spacing
\usepackage{amsmath,amssymb}
\usepackage{xcolor}         % highlighting words in Latex  \hl{}
\usepackage{soul}            % highlighting words in Latex \hl{} 
\usepackage{wasysym}
\usepackage[normalem]{ulem} % for \sout{} delete straighout line
\usepackage[utf8]{inputenc}	% prevents problems with weird characters in bibfile.
\usepackage[switch, running, displaymath, mathlines]{lineno}
\titleformat{\chapter}
       {\normalfont\fontfamily{phv}\fontsize{16}{19}\bfseries}{\thechapter}{1em}{}
\title{Binary icosahedral quasicrystals of hard spheres in spherical confinement}

\author[1,$^\|$,$\dagger$]{Da Wang}
\author[1,$^\|$]{Tonnishtha Dasgupta}
\author[1,$^\|$,$\ddagger$]{Ernest B. van der Wee}
\author[2]{Daniele Zanaga}
\author[2]{Thomas Altantzis}
\author[3]{Yaoting Wu}
\author[1]{Gabriele M. Coli}
\author[3,4]{Christopher B. Murray}
\author[2]{Sara Bals}
\author[1,*]{Marjolein Dijkstra}
\author[1,*]{Alfons van Blaaderen}

\affil[1]{Soft Condensed Matter, Debye Institute for Nanomaterials Science, Utrecht University, Princetonplein 5, 3584 CC, Utrecht, The Netherlands.}
\affil[2]{Electron Microscopy for Materials Science (EMAT), University of Antwerp, Groenenborgerlaan 171, 2020 Antwerp, Belgium.}
\affil[3]{Department of Chemistry, University of Pennsylvania, Philadelphia, PA 19104, United States.}
\affil[4]{Department of Materials Science and Engineering, University of Pennsylvania, Philadelphia, PA 19104, United States.}
\affil[$^\|$]{These authors contributed equally to this work.}
\affil[$\dagger$]{Present address: Electron Microscopy for Materials Science (EMAT), University of Antwerp, Groenenborgerlaan 171, 2020 Antwerp, Belgium.}
\affil[$\ddagger$]{Present address: Department of Physics \& Astronomy, Northwestern University, Evanston, IL 60208, United States.}
\affil[*]{e-mail: m.dijkstra@uu.nl; a.vanblaaderen@uu.nl}

\date{} %                    %% if you don't need date to appear

\begin{document}

\maketitle

%To do: check the way of supplemenatry Figs, methods, tables etc. 
% determine what Figs can go to extended data
% See Methods section for details.(https://www.nature.com/articles/s41586-018-0788-5#Sec2) ..or just use supplementary section xx (https://www.nature.com/articles/s41586-018-0770-2)

\noindent \textbf{
The influence of geometry on the local and global packing of particles is important to many fundamental and applied research themes such as the structure and stability of liquids, crystals and glasses. Here, we show by experiments and simulations that a binary mixture of hard-sphere-like particles crystallizing into the MgZn$_2$ Laves phase in bulk, spontaneously forms 3D icosahedral quasicrystals in slowly drying droplets. Moreover, the local symmetry of 70-80\% of the particles changes to that of the MgCu$_2$ Laves phase. Both of these findings are significant for photonic applications. If the stoichiometry deviates from that of the Laves phase, our experiments show that the crystallization of MgZn$_2$ is hardly affected by the spherical confinement. Our simulations show that the quasicrystals nucleate away from the spherical boundary and grow along five-fold symmetric structures. Our findings not only open the way for particle-level studies of nucleation and growth of 3D quasicrystals, but also of binary crystallization.
}

%\DA{General questions: 
%1. we have different naming of our SP through the paper: binary iQCs, iQC cluster.... should we unify the naming?
%2. through the paper, we sometimes described our particles as HS-like, sometimes HSs. I think we might also need unify the naming.
%3. Off-stochiometry SPs: we mentioned in abstract, main text and conclusion, but we do not put anything to prove our points. Shall we put 1-2 TEM images (see our Laves thesis chapter) in the SI? Claiming sth but without any evidence would definitely be questioned by reviewers.
%4. When describing our dataset, which term should we use MgCu2-like symmetry or MgCu2 symmetry? Both terms appeared in the manuscript}
%\section*{Main}
\subsection*{Introduction}

Crystallization experiments in which droplets were used have contributed significantly to our current understanding on structural aspects and the kinetics of phase transformations in condensed matter physics. 
Over half a century ago, David Turnbull \textit{et al.} showed in seminal experiments that liquids could be undercooled by as much as hundreds degrees below the freezing temperature if they were dispersed as small droplets \autocite{turnbull1956phase}. The small size of the droplets ensured the absence of heterogeneous nucleation sites, and as a consequence nucleation could only occur homogeneously. The large undercoolings demonstrated that the local structure of liquids differs substantially from that of crystals. These findings inspired Charles Frank to hypothesise that the short-range order in simple liquids has icosahedral symmetry, which is incompatible with the perfect long-range translational order of crystals \autocite{frank1952supercooling}. The icosahedral order arises locally when one maximises the density, using the convex hull, of a packing of 12 identical spheres on the surface of a central sphere with the same size \autocite{Taffs2016fivefold}. The densest packing is obtained by arranging the outer spheres on the vertices of an icosahedron, rather than by using 13-sphere subunits of face-centered cubic (FCC) and hexagonal close-packed (HCP) bulk crystals, which are known to closely pack 3D space at $\sim$74\% \autocite{steinhardt1983bond,nelson1989polytetrahedral,spaepen2000condensed}.

Long-range icosahedral order is found in icosahedral quasicrystals (iQCs), one class of quasicrystals which shows quasiperiodicity in three dimensions~\autocite{shechtman1984metallic,tsai2003review}. 
Due to their higher point group symmetry than ordinary crystals, iQCs have been demonstrated to be an ideal candidate for complete bandgap photonic crystals~\autocite{man2005experimental}. 
iQCs have been abundantly discovered in binary, ternary and multinary intermetallic compounds~\autocite{tsai2003review}. 
Polygonal quasicrystals have been found in non-atomic or soft-matter systems, such as assemblies of NCs and star polymers, whereas iQCs have not yet been found in such systems~\autocite{dotera2011quasicrystals,talapin2009quasicrystalline,ye2017quasicrystalline,Nagaoka2018}. 
Recently Engel \textit{et al.} showed by computer simulations that iQCs can be obtained by the self-assembly of one-component particles interacting \textit{via} an oscillating isotropic pair potential \autocite{engel2014ico}. Although the realisation of such a potential is non-trivial, this work suggests routes for designing iQCs in precisely tailored systems in the future~\autocite{engel2014ico}.

Colloidal suspensions are widely used as model systems to study fundamental processes such as crystallization, melting, nucleation, and the glass transition \autocite{van1995real,gasser2001real,kegel2000direct,cheng2016colloidal,fernandez2016fluids}. This is because they have a well-defined thermodynamic temperature, and consequently they display a phase behaviour similar to atoms and molecules. Additionally, they can be investigated on a single-particle level, even in concentrated systems \autocite{van1995real,gasser2001real,kegel2000direct,cheng2016colloidal,fernandez2016fluids}. 
The use of droplets has also been of interest in the soft-matter field, not only because it is a powerful method to study protein crystallization, topological defects in liquid crystal droplets, and the crystallization kinetics of colloids \autocite{cheng2016colloidal,fernandez2016fluids}, but also to fabricate clusters of particles composed of smaller particles, also termed `supraparticles’ (SPs), upon drying the droplets \autocite{wintzheimer2018supraparticles,wang2013colloidal}. 
An unexpected and recent finding in both experiments and simulations is that when up to 90,000 hard spheres (HSs) are compressed under the spherical confinement of a slowly drying emulsion droplet, the system spontaneously crystallizes into icosahedral clusters and not into the thermodynamically stable FCC bulk crystal \autocite{de2015entropy}. 
Using crystallization in spherical confinement, icosahedral clusters were obtained for a range of particle systems \autocite{Lacava2012,Yang2018binary,Wang2018magic,wang2018interplay}. % adding vogel and Wang's Nat.Commun. if comes out!
Free-energy calculations on pure HSs confined in a spherical confinement showed that clusters with an icosahedral symmetry are thermodynamically more stable than FCC clusters at packing fractions just above freezing \autocite{de2015entropy}. 
This raises the question whether spherical confinement is also able to change the equilibrium structures that form in the case of binary mixtures of spheres, and possibly also induce icosahedral symmetry. 
Here we demonstrate, by experiments and simulations, that a binary mixture of hard-sphere-like (HS-like) colloids with a size ratio of 0.78, which crystallizes into a structure analogous to the binary metallic MgZn$_2$ Laves phase in bulk, forms an iQC induced by the spherical confinement of a slowly drying droplet. 

\subsection*{Self-assembly of binary nanocrystals in spherical confinement}
\begin{figure}[!t]
\includegraphics[width=0.80\textwidth]{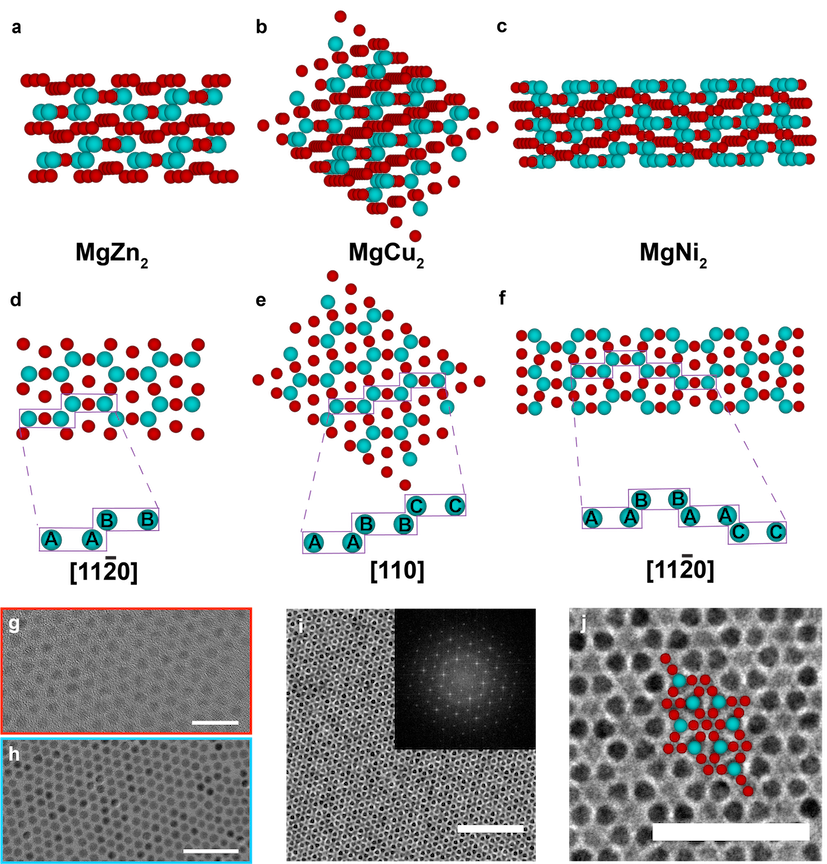}
\centering
\caption{
\textbf{Binary Laves crystal structures.}
3D views of the a) C14 (MgZn$_2$), b) C15 (MgCu$_2$) and c) C36 (MgNi$_2$) structures. 
d) C14 structure viewed along the [11$\bar{2}$0]  projection.
e) C15 structure viewed along the [110] projection.
f) C36 structure viewed along the [11$\bar{2}$0]  projection.
Transmission electron microscopy (TEM) images of g) \textit{S} CdSe nanocrystals (NCs), h) \textit{L} PbSe NCs and i-j) a self-assembled binary NC superlattice composed of CdSe and PbSe NCs showing a MgZn$_2$-like structure viewed along the [0001] projection at different magnifications. 
Fourier transform (FT) of the entire image (inset in i) with a MgZn$_2$ structure model (overlay in j) viewed along the [0001] projection, showing hexagonal symmetry. 
\textit{L} and \textit{S} spheres are coloured in cyan and red, respectively. Scale bars, g-j): 20 nm, 50 nm, 100 nm and 50 nm, respectively.
For an interactive 3D view of the three Laves phases, see Supplementary Data 1-3.
\label{Figure_1_bulk}}
\end{figure}

The Laves phases (\textit{LS$_2$}) \autocite{Laves1936,Berry1953} are three closely related structures composed of small (\textit{S}) and large (\textit{L}) spheres as schematically illustrated in Figs.~\ref{Figure_1_bulk}a-c: the hexagonal C14 MgZn$_2$ structure, the cubic C15 MgCu$_2$ structure and the hexagonal C36 MgNi$_2$ structure. The three structures can also be identified by the stacking sequence of the \textit{L} spheres as viewed along the [11$\overline{2}$0] or [110] projections of the hexagonal or cubic structures, respectively (Figs.~\ref{Figure_1_bulk}d-f) \autocite{hynninen2007self}. Specifically, the stacking of the \textit{L} sphere-dimers in the C14, C15, and C36 structures is ...\textit{AABB}..., ...\textit{AABBCC}..., and ...\textit{AABBAACC}..., respectively \autocite{hynninen2007self}. The MgZn$_2$ phase was found to be the thermodynamically stable structure for a binary mixture of \textit{L} and \textit{S} HSs with a diameter ratio in the range of 0.76-0.84, although the free-energy difference between the three Laves phases is small ($\sim$10$^{-3}$ $k_BT$ per particle at freezing) \autocite{hynninen2007self,hynninen2009stability}. 

To study the behaviour of a binary nanocrystal (NC) suspension in spherical confinement, we synthesised monodisperse CdSe NCs (\textit{S}) with a total diameter of 7.7 nm including ligands and monodisperse PbSe NCs (\textit{L}) with a total diameter of 9.9 nm including ligands (Figs.~\ref{Figure_1_bulk}g-h; Extended Data Figs. \ref{ExtendedFig_NCs_BNSL}a-b; Supplementary Section 2), yielding a binary NC mixture with a diameter ratio of 0.78. 
To investigate the bulk phase behaviour of the binary NC suspension, the mixture was allowed to self-assemble on a diethylene glycol surface according to a well-developed liquid-air interface method \autocite{Dong2010SA}. 
%We present transmission electron microscopy (TEM) images of the self-assembled binary \hl{NC} superlattice (Figs. \ref{Figure_1_bulk}i-j) showing good agreement with the [0001] projection of a MgZn$_2$ lattice (Fig.~\ref{Figure_1_bulk}i and Extended Data Fig. \ref{ExtendedFig_NCs_BNSL}c) \autocite{Wu2017}. 
The Fourier Transform (FT) (Fig. \ref{Figure_1_bulk}i inset) of the transmission electron microscopy (TEM) image of the self-assembled binary NC superlattices (Figs. \ref{Figure_1_bulk}i and Extended Data Fig. \ref{ExtendedFig_NCs_BNSL}c) shows hexagonal order, corresponding to the [0001] projection of the MgZn$_2$ structure (Fig. \ref{Figure_1_bulk}j).

%\clearpage

\begin{figure}[!t]
\includegraphics[width=0.8\textwidth]{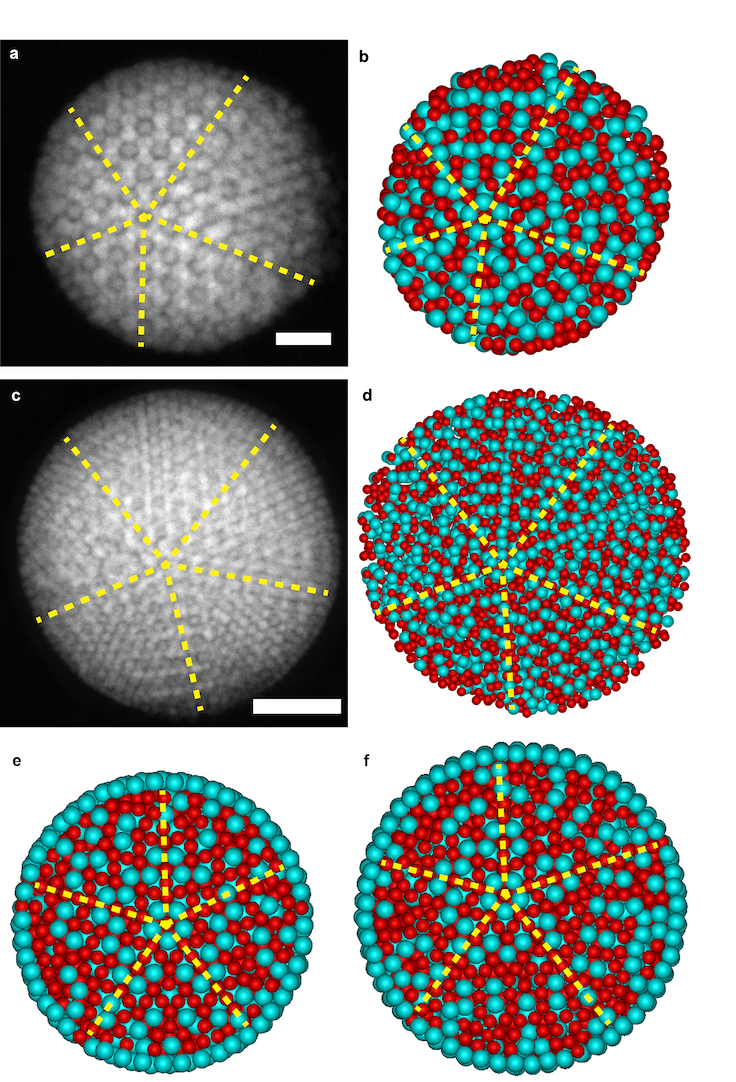}
\centering
\caption{
\textbf{Binary supraparticles (SPs) as obtained from experiments and computer simulations.}
HAADF-STEM images of two typical experimental SPs with diameters of a) 110 nm and c) 187 nm, containing CdSe/PbSe binary NCs.
b,d) Cross-section views of 3D representations of the two experimental SPs using electron tomography. 
Cross-section views of simulated SPs self-assembled from a binary mixture of e) $N_{tot}=$ 3,501 and f) $N_{tot}=$ 5,001 HS-like particles, exhibiting five-fold symmetry, similar to the experimental SPs shown in a-d.
Five-fold symmetry is visible in both the experimental and simulated SPs, as indicated by the yellow dashed lines. 
Scale bars, a): 20 nm, c) 50 nm. 
For an interactive 3D view of b and f, see Supplementary Data 4 and 5, respectively. 
For the number of particles of the experimental and simulated SPs, refer to Supplementary Tables 1 and 2. 
\label{Figure_2_SP}}
\end{figure}

Subsequently, we let the mixture of NCs with a 1:2 number ratio (\textit{L}:\textit{S}) self-assemble in slowly drying emulsion droplets \autocite{de2015entropy}, and investigated the structure of the self-assembled SPs using high angle annular dark-field scanning transmission electron microscopy (HAADF-STEM). 
We found that about 80\% of our SPs containing a few hundred up to tens of thousands of NCs show five-fold symmetry, pointing to possible icosahedral ordering (Figs. \ref{Figure_2_SP}a,c; Extended Data Fig. \ref{ExtendedFig_ExtendedFig_Exp_SPs_overview_and_tilting_v2}). 
However, since HAADF-STEM images only correspond to a 2 dimensional (2D) projection of a 3D object, we performed electron tomography \autocite{Saghi2012,Bals2014} to analyse the 3D structure of 6 SPs of different sizes (Figs. \ref{Figure_2_SP}a,c; Supplementary Videos 1 and 3; Extended Data Figs. \ref{ExtendedFig_TEM_five_fold_gallery}a-d). 
By using an advanced tomographic reconstruction algorithm \autocite{zanaga2016}, recently developed for quantitative analysis of assemblies of spherical nanoparticles, we obtained coordinates of all particles and identified both the \textit{S} and \textit{L} species in the 6 SPs (Figs. \ref{Figure_2_SP}b,d; Extended Data Figs. \ref{Extended_Fig_reconstruction} and \ref{ExtendedFig_TEM_five_fold_gallery}e-h; Supplementary Videos 2 and 4).
As will be detailed in the next section, we confirmed the presence of iQCs in all SPs which exhibited five-fold symmetry. 

It is important to note that the self-assembly in our experiments was hardly influenced by the spherical confinement when the number ratio of binary mixtures corresponded to an excess of \textit{S} spheres compared to the stoichiometry of the Laves phase. 
In these cases we found that the NCs in the SPs self-assembled into a MgZn$_2$ crystal structure as visible in 2D HAADF-STEM images, not yielding five-fold symmetry (Supplementary Fig. 1).  
%but with stacking faults, similarly as quite recently also was found for HS bulk binary crystallization of much larger colloidal HS particles analyzed by scattering \autocite{schaertl2018formation}, as the free energy differences of the three Laves as shown in Fig. \ref{Figure_1_bulk} are very similar \autocite{hynninen2009stability}. 
There are only a handful of papers on binary SPs \autocite{wintzheimer2018supraparticles,Wangpengpeng2018binary,Yang2018binary,Chen2014,Kister2016,Yang2016}, but only one recent paper by Dong \textit{et al.} presents MgZn$_2$-type binary crystals \autocite{Yang2018binary}. 
However, they concluded that a spherical confinement does \textit{not} change the structure of the binary SPs. 
Instead, we hypothesise that the difference in the effect of spherical confinement on the crystal structure of binary SPs is due to the sensitivity to the stoichiometry.

We corroborated our experimental findings with molecular dynamics (MD) simulations, using HOOMD-blue \autocite{Anderson2008,Glaser2015}, where a
binary mixture of spheres with a diameter ratio of 0.78 interacting with a HS-like potential inside a spherical confinement nucleated to form iQC clusters. 
We simulated $N_{\small tot}$ = 3,501, 5,001 and 10,002 spheres under spherical confinement (Extended Data Fig.~\ref{ExtendedFig_Simulated_Colored}) and observed iQCs for all these system sizes, thus indicating that a crossover to the MgZn$_2$ crystal phase requires a bigger system size. This expectation is in line with the observed structural crossover, observed in monodisperse HSs, from icosahedral clusters or larger surface-reconstructed icosahedral clusters to FCC ordering (which is the stable crystal phase in bulk) in system sizes with $>$ 90,000 HSs \autocite{de2015entropy}. 
The simulations were performed in a fixed spherical volume to speed up equilibration. The number densities, and
%\hl{The simulations were performed on a constant spherical volume to maximise equilibration time for core densities $\rho\sigma_\textrm{\scriptsize av, core}^{\small 3}$ between 0.867--0.945 (see Methods for definition of $\rho\sigma_\textrm{\scriptsize av, core}^{\small 3}$). 
the corresponding \emph{effective} densities for an equivalent binary HS mixture with a diameter ratio of 0.78, are presented in the Methods and Supplementary Section 5. 
The striking agreement between the experimental and simulated SPs with regard to the observation of five-fold symmetry is shown in Figs.~\ref{Figure_2_SP} and \ref{Figure_3_Diff}.
%These densities are equivalent to an \emph{effective density} range 0.966--1.138 for a binary hard-sphere mixture of diameter ratio $q$ = 0.78~(see Supplementary Section 5 for effective hard-sphere density calculation).}

%We speculate that this is caused by stochiometry effects, which are thus quite important \DA{for binary crystallization in spherical confinement.}

\subsection*{Fourier-space analysis of icosahedral quasicrystals}
\begin{figure}[!t]
\includegraphics[width=0.8\textwidth]{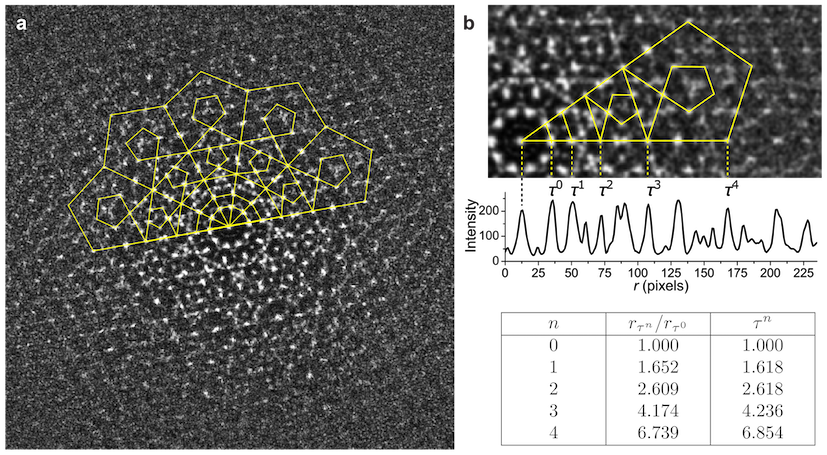}
\centering
\caption{
\textbf{Calculated diffraction pattern of a simulated binary icosahedral SP.}
a) The calculated diffraction pattern of a simulated binary SP containing $N_{\small tot}$ = 5,001 particles, showing five-fold orientational symmetry. 
%The FT is calculated on time-averaged coordinates of particles belonging to the iQC cluster identified using BOPs (Supplementary Section 5).
The yellow lines highlight the presence of inflating pentagons in the calculated diffraction pattern, which is a direct consequence of the non-crystallographic five-fold orientational symmetry, depicting quasiperiodicity.
b) One of the ten domains in the calculated diffraction pattern showing the inflating pentagon sequence. The intensity profile along the horizontal yellow line features peaks whose positions follow the $\tau^n$ sequence, as listed in the table. 
The bare calculated diffraction pattern and a full overlay can be found in Supplementary Fig. 3.
\label{Figure_3_Diff}}
\end{figure}

Inspired by the beautiful work of Shechtman \textit{et al.} \autocite{shechtman1984metallic} that led to the discovery of iQCs, we calculated diffraction patterns from our real-space data.  %Shechtman1984 paper did not mention 10-fold! Finger prints of iQCs are six 5F rotation axes, ten 3F axes and fifteen 2F axes. Also 5F, 3F and 2F appeared in Glotzer's Nature Mater. https://www.nature.com/articles/nmat4152 
iQCs have 3D long-range order, but are \textit{not} periodic and therefore display intensity peaks in their diffraction patterns, with symmetries that are not compatible with a periodic repeating unit, such as five-fold rotational symmetry (Fig. \ref{Figure_3_Diff}). 
We first calculated diffraction patterns of an experimental SP, which were determined by calculating the FTs of 2D projections of the coordinates (Supplementary Fig. 2). Next, we calculated
diffraction patterns of data sets corresponding to a simulated SP ($N_{tot}$ = 5,001; Fig. \ref{Figure_3_Diff}) using the same method, but we averaged
the coordinates over a relatively long time, and only selected the particles belonging to the iQC cluster according to criteria based on
bond-orientational order parameters (BOPs) (Methods and Supplementary Section 4).  
The sharp Bragg peaks exhibit five-fold (Figs. \ref{Figure_3_Diff}a-b) as well as two-fold and three-fold orientational symmetries (Extended Data Fig. \ref{ExtendedFig_2-3-fold_symmetries_FFT}), which was also observed for simulated one-component iQCs \autocite{engel2014ico}. As the averaging of the coordinates is not applicable to the experimental data set, the calculated diffraction pattern does not show many higher order Bragg peaks (Supplementary Fig. 2). 
However, as will be detailed in the next section, the SPs obtained from experiments and simulations were found to have the same structure in real space. 

The calculated diffraction pattern in Fig. \ref{Figure_3_Diff} shows inflated pentagons following a $\tau^n$ sequence, where $\tau$ = $(1+\sqrt{5})/2$ is the irrational golden mean. This so-called $\tau^n$ sequence is a direct consequence of the non-crystallographic icosahedral symmetry and is characteristic of iQCs~\autocite{tsai2003review}. 
The perfect agreement of the positions of the Bragg peaks in our data along a projection vector as shown in Fig. \ref{Figure_3_Diff}b, with the $\tau^n$ sequence, signifies that our SP structure is indeed an iQC. For instance, multiple-twinned crystals would show diffraction patterns similar to those of iQCs but the Bragg peaks would display a twinned substructure in the diffraction pattern, which we do not see.

\begin{figure}[!t]
\includegraphics[width=0.8\textwidth]{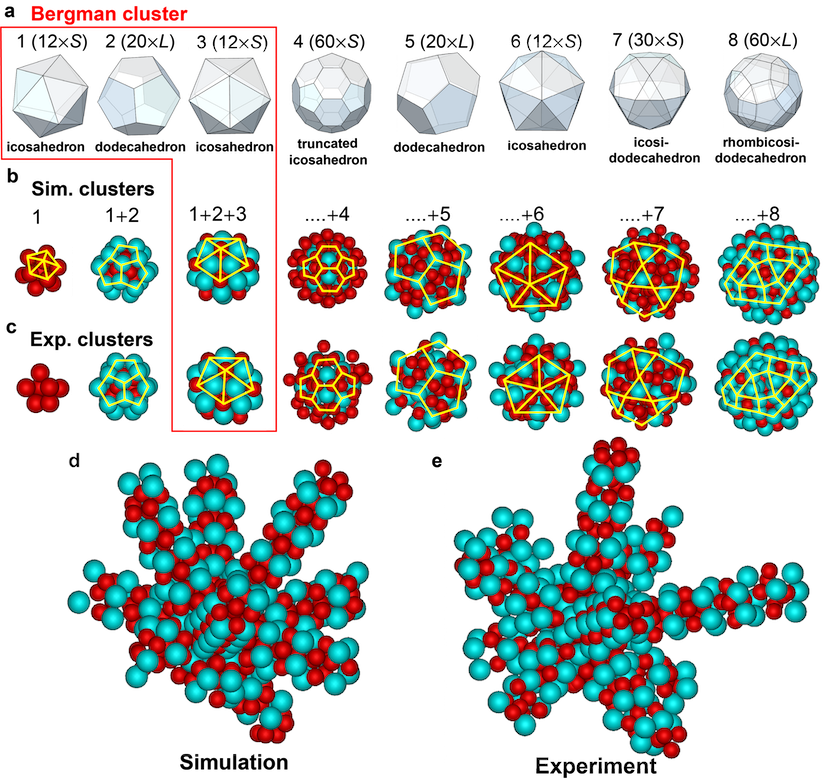}
\centering
\caption{
\textbf{Real-space structure of the iQC cluster.}
iQC clusters were found to contain concentric shells around the center of symmetry with geometries corresponding to Platonic and Archimedean solids.
a) The regular polygons describing the shells up to the 8$^\textrm{\small th}$ shell and the number of particles (of $L$ or $S$ species) forming them.
Corresponding regular shells as found in b) simulations and c) experiments, where \textit{L} and \textit{S} spheres are coloured in cyan and red, respectively. The yellow lines denote the regular polygons corresponding to the surface tiling of the geometric solids.
The 44-particle cluster up to the 3$^\textrm{\small rd}$ shell, as delineated by the red box, corresponds to the Bergman cluster.
Along the five-fold symmetry axes of the iQC, pentagonal tubes composed of alternating pentagons of $L$ and $S$ particles were found in both d) simulations and e) experiments, all converging at the center of a Bergman cluster. Interactive 3D views of b-e) can be found in Supplementary Data 6-9.
\label{Figure_4_clusters}}
\end{figure}

\subsection*{Real-space analysis of icosahedral quasicrystals}
Fundamental clusters that are often used in the classification of iQCs are the Mackay cluster \autocite{mackay1962dense,kuo2002mackay}, the Bergman cluster \autocite{bergman1957crystal,steurer2008fascinating}, and the Tsai cluster \autocite{Tsai2000Tsaicluster,maezawa2004icosahedral}, which consist of concentric shells with regular polygonic shapes. 
%\hl{Upon inspecting our experimental and simulated iQC clusters in more detail}, we observed that the SPs indeed contained such concentric shells of either \textit{S} or \textit{L} particles around the center of symmetry (Figs.~\ref{Figure_4_clusters}a-c). 
We found that both our experimental and simulated SPs contained such concentric shells of either \textit{S} or \textit{L} particles around a center of symmetry (Figs.~\ref{Figure_4_clusters}a-c).
We found that these clusters can be classified as Bergman clusters and positioned not at the center of the spherical confinement.
The experimental and simulated SPs show striking structural similarity (Fig.~\ref{Figure_4_clusters}).
iQCs have been found in many binary and ternary intermetallic compounds \autocite{tsai2003review,steurer2008fascinating}. To the best of our knowledge, it is not yet known that equilibrium iQCs can form from HS-like particles induced by a spherical confinement. %, nor have binary iQCs been observed which contain Bergman clusters. not true, see: Takakura et al 2006 Nat Materials

In addition, we analysed the local structural environment of the particles in the SPs using a BOP analysis \autocite{lechner2008accurate} on the \textit{L} particles only (Supplementary Section 3). 
Interestingly, the BOPs of the \textit{L} particles in the iQC cluster coincide with the BOP signatures of the \textit{L} particles in both equilibrated MgZn$_2$ and MgCu$_2$ Laves phases in bulk. % Refer to the BOP section in SI and fig. EvdW prepared (bop)
We present typical configurations of experimental and simulated iQC clusters, in which the \textit{L} particles are coloured according to their different BOP values (Extended Data Figs. \ref{ExtendedFig_BOP}a-d). 
%The MgCu$_2$-like \textit{L} particles are coloured in blue, and MgZn$_2$-like \textit{L} particles are shown in red. 
We observe that the iQC clusters consist of twenty tetrahedral domains (or wedges) that predominantly consist of particles with a MgCu$_2$-like symmetry, whereas the \textit{shared faces} of adjacent tetrahedral domains are MgZn$_2$-like (Extended Data Figs. \ref{ExtendedFig_BOP}b,d). 
Based on single-particle systems, where the tetrahedral domains of the icosahedral clusters were found to be (cubic) FCC-like \autocite{de2015entropy}, it is not surprising that the wedges of our icosahedral SP structures follow a cubic MgCu$_2$-like symmetry. 
Therefore, effectively, the hexagonal MgZn$_2$ Laves phase which is obtained in bulk has been transformed into a SP comprised of domains with a local symmetry similar to the cubic MgCu$_2$ Laves phase. 
We further estimated the fraction of MgCu$_2$-like particles in experimental and simulated SPs, for different SP sizes. 
We see that the spherical confinement forces the local symmetry of 70-80\% of the particles in the SP to resemble that of the MgCu$_2$ Laves phase (Extended Data Fig. \ref{ExtendedFig_BOP}e).
In addition, the \textit{edges} of five neighbouring tetrahedral domains form a pentagonal tube (or icosahedral chain~\autocite{yang2018precipitation}) that originates from the Bergman cluster, runs through the whole SP, and ends at one of the twelve vertices with five-fold symmetry on the surface of the SP (Figs. \ref{Figure_4_clusters}d-e).
%We show \hl{the} pentagonal tubes present in our experimental (Fig. \ref{Figure_4_clusters}d) and simulated SPs (Fig. \ref{Figure_4_clusters}e). 
We remark that the pentagonal tubes differ in length as all the Bergman clusters that were found were eccentric with respect to the spherical confinement. 
We thus find that a binary mixture of HS-like particles, that crystallizes into the MgZn$_2$ Laves phase in bulk, forms an iQC composed of tetrahedral domains with MgCu$_2$-like symmetry induced by a spherical confinement. 

\subsection*{Nucleation and growth of icosahedral quasicrystals}
\begin{figure}[!t]
\includegraphics[width=0.8\textwidth]{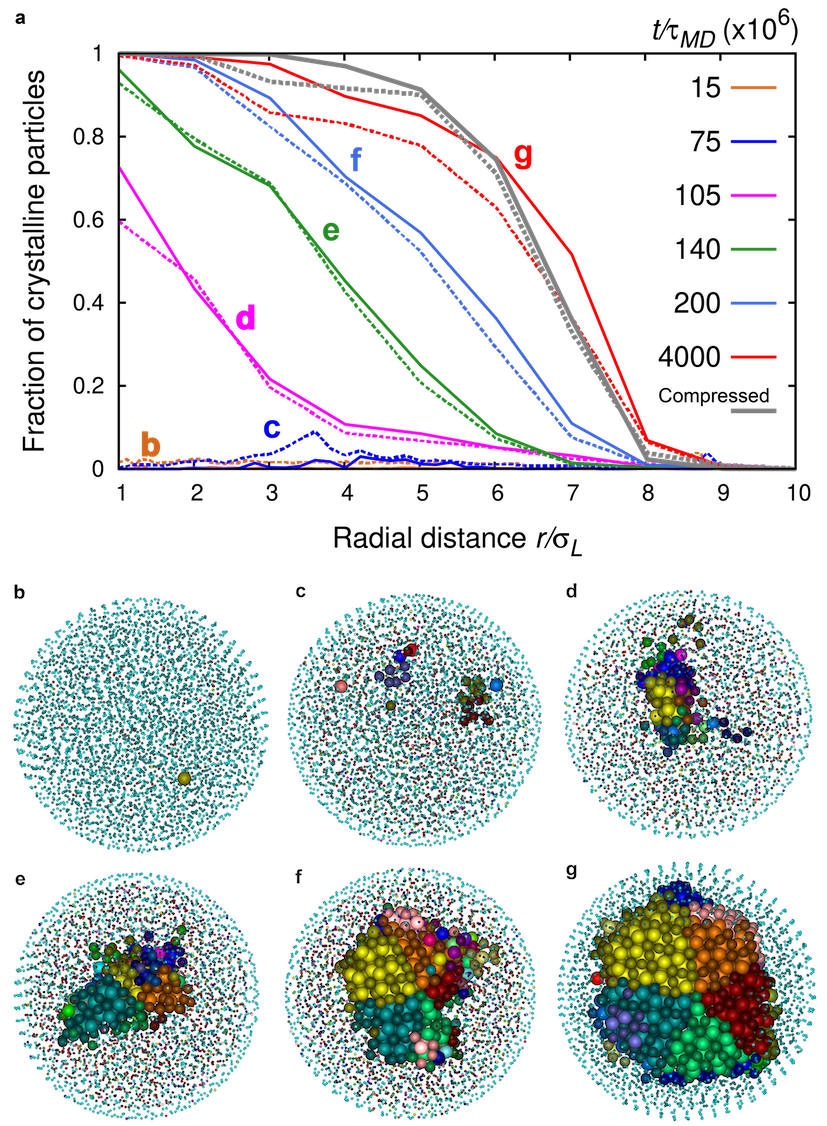}
\centering
\caption{
\textbf{Nucleation and growth of the iQC cluster in simulations.} a) The fraction of crystalline particles as a function of the radial distance from the center of the SP with $N_{tot}$ = 5,001 particles at different MD times. Crystalline particles belonging to the iQC cluster are identified using BOPs. The solid lines denote the crystalline fraction of $L$ spheres and the dashed lines represent that of the $S$ species. The crystalline fraction is calculated over the total number of particles of the same species and averaged over 1,000 configurations. The profile shows that the crystal nucleates at an off-centered position and grows radially outward with time. b-g): Typical configurations corresponding to the crystallization profiles shown in a), the labels (b-g) are shown in a) next to and in the same colour as the corresponding crystallization profile, for clarity. The different crystalline domains are shown in different colours. For crystalline and domain classification, see Supplementary Section 4. The $S$ spheres are coloured in a darker shade of the same colour as the $L$ spheres. Particles not belonging to the iQC cluster are reduced in size for visual clarity.  
%$\rho\sigma_\textrm{\scriptsize av}^{\small 3}$ ($\rho\sigma_\textrm{\scriptsize HS, eff}^{\small 3}$) = 0.809 (0.974). 
A video of the nucleation of the iQC cluster is shown in Supplementary Video 5.
\label{Figure_5_nucleation}}
\end{figure}

Now that we have established that the structure of our SPs corresponds to a binary iQC cluster, we investigated one of the many open questions on iQCs \autocite{Steurer2018}, namely, how do iQCs grow? Does our binary iQC nucleate from a Bergman cluster and grow shell-by-shell?  
Or does crystallization start at the spherical interface similar to the formation of icosahedral clusters from single-component HSs in spherical confinement \autocite{de2015entropy}? 
In order to study the nucleation and growth of iQCs from a supersaturated binary fluid phase, we used a BOP-based criterion to identify and track the crystal nucleation (Supplementary Section 4) in our simulations inside a spherical confinement at fixed volume.
%with a number density $\rho\sigma_\textrm{\scriptsize av}^{\small 3}$ = 0.809. The effective density $\rho\sigma_\textrm{\scriptsize HS, eff}^{\small 3}$ of an equivalent binary hard-sphere (BHS) mixture  with a diameter ratio 0.78 is 0.974 (see Supplementary Section 5 for details).  
In Fig. \ref{Figure_5_nucleation}a, we plot the fraction of crystalline particles as a function of the radial distance from the center of a SP consisting of $N_{tot}$ = 5,001 particles at different times along with corresponding typical configurations (Figs. \ref{Figure_5_nucleation}b-g). 
The solid lines denote the crystalline fraction of \textit{L} spheres while the dashed lines represent the crystalline fraction of the \textit{S} spheres. 
The evolution of the structure within the SP shows all the characteristics of nucleation and growth (Figs. \ref{Figure_5_nucleation}b-g; Supplementary Video 5). 
We find that the system remains for a long time in the supersaturated fluid phase, in which small crystalline clusters form and melt in subsequent frames in the spherical confinement (Fig. \ref{Figure_5_nucleation}c). 
At longer times, a single primary cluster nucleates at an off-centered position and starts to grow (Fig. \ref{Figure_5_nucleation}d). 
The nucleus diffuses slightly towards the center of the spherical confinement and grows into the surrounding fluid with time (Figs. \ref{Figure_5_nucleation}e-g). 
This can also be appreciated from Fig. \ref{Figure_5_nucleation}a that shows that the crystallization front has grown radially outward with time. 
The long period that the system stayed in a supersaturated fluid phase and the induction time before a single cluster starts to grow are typical for nucleation. 
We therefore conclude that the iQC cluster nucleates away from the spherical boundary.
In addition, we find that first a pentagonal tube is formed from which the structure grows out. The growth is therefore not initiated by the Bergman cluster.

An intriguing question is why a mixture that crystallizes into the MgZn$_2$ Laves phase in bulk forms an iQC cluster under spherical confinement. 
As the iQC cluster forms \textit{via} nucleation away from the spherical confinement, the curvature can only be communicated \textit{via} the supersaturated fluid phase. 
We measured the radial density profiles of the $L$ and $S$ species in the SP at different times in our simulations and found that the density profiles display pronounced layering of particles in concentric shells, which lends strong support to the formation of the iQC being driven by the structural correlations and frustrations induced by the spherical curvature in the highly structured fluid phase (Extended Data Fig. \ref{ExtendedFig_layering_and_correlations}). 
%From the pair correlation functions of the equilibrated supersaturated fluid, it can be seen that these structural correlations persist over long length scales  (Extended Data Fig. \ref{ExtendedFig_layering_and_correlations}b).

\subsection*{Implications for photonic crystal research}
Our finding that iQCs can self-assemble from a binary mixture of colloidal HS-like particles in spherical confinement is also interesting for photonic applications. 
Experiments on 3D printed structures in plastic have already demonstrated that iQCs possess large and overlapping stop bands for microwave radiation \autocite{man2005experimental}.
Moreover, our results show that the local symmetry of 70-80\% of the particles has been changed to MgCu$_2$-like, where the MgCu$_2$ is composed of a cubic diamond lattice of the \textit{L} spheres and a pyrochlore lattice of the \textit{S} spheres that both have a photonic band gap for low index contrasts \autocite{hynninen2007self,hynninen2009stability}. 
Although it still has to be determined theoretically which structure, either the iQC cluster or the wedges with MgCu$_2$-like symmetry, is best suited for opening up a photonic band gap at low refractive index contrast \autocite{man2005experimental}, it is almost certain that if hard particles with a large index contrast such as titania are used, for which the optical properties on the single particle level can be tuned in great detail \autocite{jannasch2012nanonewton}, a photonic band gap can be realised for visible wavelengths. 
% old: Although it has recently become possible to grow MgCu$_2$ colloidal crystals using DNA interactions \autocite{ducrot2017colloidal}, 
Although it has recently become possible to grow photonic crystals with diamond sub-structures using DNA interactions \autocite{ducrot2017colloidal,wang2017colloidal},
the procedures are complicated and the crystals and index contrasts are not yet good enough to open up photonic band gaps. 
Crystallization of HS-like particles with a large refractive index contrast provides a much easier route to obtain a band gap for visible light. 
Moreover, as our results have been obtained for HS-like particles, it will be fairly straightforward to extend these results to particles with a size closer to the wavelengths of visible light. 

\subsection*{Conclusion}
Our findings in this paper have strengthened the connection already made in previous work between icosahedral order and frustration brought about by curved space \autocite{de2015entropy,guerra2018freezing,sadoc1981use}. The spherical shape of the SPs also allowed a quantitative real-space analysis of the iQCs that are formed when binary mixtures of HS-like particles with a size ratio of 0.78 are made to crystallize at a stoichiometry corresponding to a MgZn$_2$ Laves phase. The iQCs were found to consist of Bergman clusters with additional concentric shells of particles. In addition, we found that the local symmetry of the particles in the tetrahedral domains of the icosahedral cluster transformed to a MgCu$_2$-like symmetry which is, next to an iQC itself, an important and much sought after structure in photonic crystal research. Furthermore, we studied the evolution of the iQC in computer simulations, and tracked the nucleation and growth mechanism by identifying the particles belonging to the iQC cluster using BOPs. 
We observed that the iQC clusters nucleate away from the spherical boundary and that the iQC growth is induced by the pronounced layering of the highly-structured fluid at the spherical boundary.  
As recently HS-like MgZn$_2$ colloidal crystals have been realised \autocite{schaertl2018formation}, it is likely that by having such systems self-assemble inside a spherical confinement interesting colloidal photonic iQCs can now be realised as well.
Finally, our experimental finding that the effect of the spherical confinement can be circumvented by tuning the composition to off-stoichiometric values, is likely transferrable to other binary crystal structures in the rich and complex binary phase behaviour. 
This means that the droplet platform as explored here, but using larger colloidal particles that can be imaged in 3D by light nanoscopy, can also be used to obtain the first experimental single particle level information on binary nucleation and growth. 

\subsection*{Methods}

{\footnotesize
\noindent \textbf{Nanocrystal (NC) syntheses} 7.7 nm CdSe NCs \autocite{Pietryga2008} (a total polydispersity of 2\% by counting 130 CdSe NCs; 5.2 nm core diameter) and 9.9 nm PbSe NCs \autocite{steckel2006} (a total polydispersity of 1\% by counting 170 PbSe NCs; 7.6 nm core diameter) were synthesised with minor modifications according to literature methods. 
This yields a binary NC mixture with a diameter ratio of 0.78. 
The NC syntheses were carried out in a nitrogen atmosphere using standard Schlenk line techniques. 
As-synthesised NCs were purified by isopropanol and were dispersed in cyclohexane with desired weight concentrations. All NCs used in our current work are washed well before self-assembly according to their original experimental protocols. Thus one can assume the amount of free ligands to be negligible after proper washing such that they do not play a significant role during the self-assembly. 
Our building blocks (cores and ligands) are similar to those used in the work of Evers \textit{et al.} \autocite{Evers2010}, we therefore describe our experimental system with a hard-sphere-like (HS-like) potential.
Detailed information relating to the syntheses can be found in Supplementary Sections 1-2.

\noindent \textbf{Experimental self-assembly of binary NCs in bulk}
Binary NC superlattices were formed at a liquid-air interface \autocite{Dong2010SA} in a glovebox that maintains oxygen- and moisture-free conditions. To prepare a typical MgZn$_2$-type binary NC superlattice, 
5.2 nm CdSe and 7.7 nm PbSe NCs were separately dispersed in \textit{n}-hexane at a concentration of 10 mg/mL. The two samples were mixed ($V_{CdSe}/V_{PbSe}\approx 0.9$) to give the right number ratio of NCs and form a high-quality MgZn$_2$-type binary NC superlattice film. Then \SI{10} {\micro L} of the mixture was dropcast onto the surface of diethylene glycol in a square Teflon well. A glass slide was placed to cover the well and reduce the evaporation rate of hexane. After 30 minutes, a solid film was obtained on the liquid-air interface. %Typical self-assembled binary NC superlattices can be found in Figs. \ref{Figure_1_bulk}i,j and Extended Data Fig. \ref{ExtendedFig_NCs_BNSL}c. 

\noindent \textbf{Experimental self-assembly of binary NCs in spherical confinement}
For a typical icosahedral quasicrystal assembled from binary NCs in spherical confinement experiment, 6.5 mg of PbSe NCs and 5.85 mg of CdSe NCs
were dispersed in 1.0 mL of cyclohexane and added to a mixture of \SI{400}{mg} of
dextran and 70 mg of sodium dodecyl
sulfate (SDS) in \SI{10}{mL} of de-ionized (DI) H$_2$O. % Millipore direct-Q will be mentioned in the SI instead of here.
The resulting emulsion was
agitated by shear with a shear rate of $1.56\times10^{5}$ $s^{-1}$, using a Couette
rotor-stator device (gap spacing 0.100 mm) following the procedure and home-built equipment
described by Mason and Bibette \autocite{mason1997}. The emulsion was then evaporated at room temperature (RT), while sedimentation was prevented by mixing the emulsion using a VWR VV3 vortex mixer for 48 hours. The resulting supraparticles (SPs) suspension was purified by centrifugation with a speed of 2,500 rpm for 15 minutes using an Eppendorf 5415C centrifuge, followed by re-dispersing in DI H$_2$O. The aforementioned procedure was repeated twice. 
%Overview of self-assembled \hl{supraparticles (SPs)} can be found in Extended Data Figs. \ref{ExtendedFig_ExtendedFig_Exp_SPs_overview_and_tilting_v2}a. 
The experimental procedure of a typical binary NC self-assembly with an off-stochiometry is similar to that for obtaining the icosahedral quasicrystal, except that 6.5 mg PbSe and 6.5 mg CdSe NCs were dispersed in 1.0 mL of cyclohexane and added to a mixture of \SI{400}{mg} of dextran and 70 mg of SDS in \SI{10}{mL} of de-ionized (DI) H$_2$O.
%(Supplementary Fig. 1). 
 
\noindent \textbf{Electron microscopy (EM) sample preparation and 2D EM measurements}
To prepare a sample for conventional EM imaging and electron tomography analysis, \SI{3} {\micro L} of the SPs
suspension in DI H$_2$O was deposited on a Quantifoil (2/2, 200 mesh) copper
grid and plunge-frozen in liquid ethane using a Vitrobot Mark2 plunge freezer
at temperatures around 90 K. The sample was then freeze-dried over a period of 8
hours under vacuum at 177 K and subsequently allowed to warm to RT prior to
EM analysis.

Conventional transmission electron microscopy (TEM), 2D high angle annular dark-field-scanning transmission electron microscopy (HAADF-STEM) and 2D energy dispersive X-Ray (EDX) spectroscopy chemical mapping measurements were performed on a FEI-Talos F200X electron microscope, equipped with a high-brightness field emission gun (X-FEG) and operated at 200 kV.

Images and elemental EDX maps were acquired using Bruker Esprit analytical and imaging software in HAADF-STEM mode. Elemental EDX maps of 418$\times$416 pixels were acquired with a 15-minute acquisition time to get a good signal-to-noise ratio, as shown in Supplementary Fig. 1.

\noindent \textbf{Electron tomography}
Electron tomography experiments were performed using an aberration-corrected `cubed' FEI-Titan electron microscope, operated at 300 kV in HAADF-STEM mode. 
Six electron tomography series of self-assembled binary SPs of different sizes were acquired using a Fischione model 2020 single-tilt tomography holder. 
For the SPs with a diameter of 80 nm, 110 nm, 132 nm, 167 nm, 170 nm and 187 nm, tilt series were acquired over the following tilt ranges:  
from -\SI{74}{\degree} to +\SI{74}{\degree}, 
from -\SI{76}{\degree} to +\SI{76}{\degree}, 
from -\SI{74}{\degree} to +\SI{74}{\degree}, 
from -\SI{74}{\degree} to +\SI{76}{\degree}, 
from -\SI{76}{\degree} to +\SI{72}{\degree} and 
from -\SI{66}{\degree} to +\SI{74}{\degree}, respectively. 
The increments in all experiments were set to be \SI{2}{\degree}.

\noindent \textbf{Tomographic reconstruction}
Alignment of the tilt series was performed using cross-correlation based routines \autocite{Guizar-Sicairos2008} developed in MATLAB. 
Tomographic reconstructions were obtained using the Simultaneous Reconstruction Technique (SIRT) \autocite{gilbert1972}, as implemented in the ASTRA toolbox, and according to the approach explained in Ref.~\cite{zanaga2016}.

The SIRT algorithm was used to obtain a preliminary reconstruction of the tilt series yielding morphological information about the 3D shape of the SP. However, in the presence of close packed particles, conventional SIRT is not sufficient to achieve the reconstruction resolution necessary to segment individual particles and obtain structural information. In order to investigate this in more detail, the Sparse Sphere Reconstruction (SSR) algorithm \autocite{zanaga2016} was used to obtain the coordinates of the particles in the assemblies. The SSR technique has been developed specifically for this purpose. The main idea behind the method is to leverage prior-knowledge in the reconstruction process in order to overcome resolution limitations of SIRT when reconstructing assemblies of close-packed particles. The assumption used is that the particles in the assembly are monodisperse spheres of known size. 

In our case the assemblies present two types of particles. However, the particles differ only slightly in size and density, and at the resolutions used for the reconstructions, the difference in size between the two species was less than 2 pixels. Therefore the SSR algorithm could be applied using the approximation of monodisperse particles, and only one size (estimated from the SIRT reconstructions as the size of the smaller species) was used as prior. Once the coordinates of the particles were obtained from the SSR reconstructions, the SIRT reconstruction was used to differentiate the two species which were labelled \textit{S} species (CdSe NC) and \textit{L} species (PbSe NC).

To distinguish them, a spherical neighbourhood around each center coordinate and with a radius equivalent to the one of the smaller species (CdSe NCs) was extracted from the SIRT volume. For each point, an average gray value was calculated for these neighbourhoods. Since tilt series were acquired in HAADF-STEM mode, the different atomic numbers of the cations create a different contrast in the SIRT volumes, which yields two different intensity distributions. %(see Supplementary Fig. 4 of the 110 nm binary SP). 
The intensity distributions were fitted with two Gaussian distributions, and where the intersection of the two curves was taken as the threshold value to distinguish the two species (Supplementary Fig. 4 of the 110 nm binary SP). 
The overlap of the two curves gives an estimate of the percentage of particles that might be misidentified (\textit{e.g.} $\sim 1\%$ for the 110 nm SP). The 3D representations of all 6 experimental binary SPs from tomographic reconstruction can be found in Figs. \ref{Figure_2_SP}b,d and Extended Data Fig. \ref{ExtendedFig_TEM_five_fold_gallery}.

In order to confirm the validity of the results of the SSR reconstructions, manual segmentation was performed on the SIRT reconstructed volumes of two binary SPs with diameters of 110 nm and 170 nm, respectively, as shown in Extended Data Fig. \ref{Extended_Fig_reconstruction}.
Due to the smaller size of the CdSe NCs and the lower contrast that exhibited in the SIRT reconstructions compared to the PbSe NCs, manual segmentation of CdSe NCs is time-consuming and far from straightforward. 
Therefore, we implemented the approach only on the $L$ PbSe NCs (Extended Data Fig. \ref{Extended_Fig_reconstruction}).
2,825 and 990 PbSe NCs were detected from the 170 nm SP and 110 nm SP, respectively. 
The number of PbSe NCs obtained from the SSR reconstructed volumes of the 170 nm and 110 nm SP is 2,814 and 982, respectively. 
The deviation of the number of the PbSe NCs that obtained from SSR and SIRT is only 0.39\% and 0.8\% for the 170 nm SP and 110 nm SP, respectively.

\noindent \textbf{Computer simulations}
We study the spontaneous nucleation of the binary iQC clusters by performing molecular dynamics (MD) simulations of a binary mixture of particles inside a spherical confinement, interacting \textit{via} the Weeks-Chandler-Andersen (WCA) pair potential
\begin{linenomath*}
\begin{align*}
\phi_{\alpha\beta} \left(r_{ij}\right) &= 4 \epsilon_{\alpha\beta} \left\lbrack \left(\frac{\sigma_{\alpha\beta}}{r_{ij}} \right)^{12} - \left(\frac{\sigma_{\alpha\beta}}{r_{ij}} \right)^6 + \frac{1}{4} \right\rbrack & r_{ij}<2^{1/6}\sigma_{\alpha\beta}\nonumber\\
&= 0 &r_{ij}\geq2^{1/6}\sigma_{\alpha\beta}, 
\label{wca}
\end{align*}
\end{linenomath*}
where $i$,$j$ are the interacting particles, $\alpha$,$\beta$ denote the particle type (\emph{i.e.} $L$,$S$). For our binary mixture, interaction strength $\epsilon_{LL} = \epsilon_{SS} = \epsilon_{LS} = \epsilon$, $\sigma_{LL} \equiv \sigma_{L}$, $\sigma_{SS} \equiv \sigma_{S}$,  diameter ratio $q = \sigma_{S}/\sigma_{L}$ = 0.78 with $\sigma_{LS} = \left(\sigma_{L} + \sigma_{S}\right)/2$, and reduced temperature $T^*=k_BT / \epsilon$ = 0.2. The WCA potential is short-ranged, steeply repulsive and reduces to the hard sphere pair potential 
%with a diameter of $2^{1/6}\sigma_{\alpha\beta}$ 
in the limit of $T^*\rightarrow 0$. The interaction potential of the particles with the spherical wall is also a repulsive WCA potential with $\epsilon_{w} = \epsilon$ and $\sigma_{w} = \sigma_{L}$. The simulations are performed using HOOMD-blue (Highly Optimized Object-oriented Many-particle Dynamics) \autocite{Anderson2008,Glaser2015} in the canonical $NVT$ ensemble, where the total number of particles $N_{tot}$ = $N_L+N_S$, temperature $T$, and the volume $V$ are fixed. The temperature is kept constant \textit{via} the Martyna-Tobias-Klein (MTK) \autocite{martyna1994constant} integration of the equations of motion based on the Nos\'e-Hoover thermostat \autocite{nose1984unified,hoover1985canonical}, with coupling constant $\tau_{\tiny T}$ = 1.0. The time step is set to $\Delta t = 0.01\tau_{\scriptsize MD}$ where $\tau_{\scriptsize MD} = \sigma_{\scriptsize L}\sqrt{m/\epsilon}$ is the MD time unit. In order to ensure sufficient time for equilibration, the simulations are performed in a constant spherical volume. The iQC nucleation and growth is tracked using a BOP-based cluster criterion as described in the Methods and Supplementary Section 4. 
The crystallized SPs are then subjected to a fast compression up to very high densities 
%($\rho\sigma_\textrm{\scriptsize av}^{\small 3} = \frac{3N_{tot}\sigma_{L}^3}{4\pi R_{\small SC}^3}\left(x_L+\left(1-x_L\right)q^3\right) \in$ 0.941--0.957), 
%0.957 (3501),0.945 (5001),0.941 (10002)
so as to arrest the positions of the spheres in the SP. The compressed SP structures are subsequently analysed using BOPs on \textit{L}-\textit{L} species correlations, as described in Supplementary Section 3. For a close comparison with the experiments, where five-fold SPs were only observed for compositions close to $LS_2$, we modulate the composition of our binary mixtures such that the number fraction of \textit{L} particles $x_{\small L} = \frac{N_{\small L}}{N_{\small L} + N_{\small S}}$ is $\sim$ 0.3 in the `core'. 
The \emph{core} comprises of the SP \emph{excluding} those particles positioned at a radial distance $> R_{\small SC} - 2\sigma_{\small L}$ from the center of the SP, where $R_{\small SC}$ denotes the radius of the spherical confinement. 
The effective number of particles in the core is denoted by $N$ while the total number of particles in the SP is denoted by $N_{\small tot}$. In this work, we simulate $N_{\small tot}$ = 3,501, 5,001 and 10,002 spheres, which correspond to an effective number of $N$ = 1,990, 3,060 and 6,870 particles in the core of the SP respectively. 
The supersaturated reduced densities $\rho\sigma_\textrm{\scriptsize av}^{\small 3}$ (= $\frac{3N_{tot}\sigma_{L}^3}{4\pi R_{\small SC}^3}\left(x_L+\left(1-x_L\right)q^3\right)$) for which crystallization is observed in our constant volume simulations lie between 0.802--0.831. These densities are equivalent to effective packing fractions $\eta_\textrm{\tiny HS, eff}$ in the range 0.506--0.524 for a binary HS mixture with a diameter ratio $q$ = 0.78. The effective packing fraction for the binary HS mixture is calculated using an \emph{effective diameter}, which can be estimated from WCA perturbation theory (see Supplementary Section 5 for details). 

\noindent \textbf{Calculated diffraction pattern} 
The calculated diffraction pattern of the simulated 5,001-particle SP was calculated from the coordinates of particles belonging to the iQC cluster after the following treatment:
(1) the compressed SP was subjected to an equilibration cycle ($t/\tau_{MD}$ = 10$^6$, $\Delta t = 10^{-3}\tau_{\scriptsize MD}$) and the particle coordinates were averaged over 10,000 configurations, and (2) the time-averaged coordinates were subsequently run through a cluster criterion based on BOPs, which distinguishes the particles belonging to the iQC cluster from particles with a different local order, as determined by $d_{6,\alpha\beta}(i,j)$ correlations of the particles with their neighbours. Here, $\alpha$,$\beta$ denotes the type of species (\emph{i.e.} $L$,$S$) and $i$,$j$ represents the index of the particle and its neighbour respectively. 
Next, the coordinates of these particles are projected on a 2D plane and the diffraction pattern is obtained by calculating the FT of the resulting projection.
The intensity profile in Fig. \ref{Figure_3_Diff}b was filtered with a 1 pixel Gaussian blur to suppress noise. Further details on the BOP criteria are provided in Supplementary Section 4.

%\printbibliography

\subsection*{Acknowledgements} 
G. A. Blab is gratefully acknowledged for 3D printing numerous truncated tetrahedra which increased our understanding of the connection between the iQC and Laves phase structures. N. Tasios is sincerely thanked for providing the code for the diffraction pattern calculation. 
M. Hermes is sincerely thanked for providing interactive views of the structures in this work. The authors thank G. van Tendeloo, S. Dussi, L. Filion, E. Boattini, S. Paliwal, N. Tasios, B. van der Meer and I. Lobato for fruitful discussions. D.W., E.B.v.d.W. and A.v.B. acknowledge partial financial support from the European Research Council under the European Union's Seventh Framework Programme (FP-2007–2013)/ERC Advanced Grant Agreement 291667 HierarSACol. T.D. and M.D. acknowledge financial support from the Industrial Partnership Programme, ``Computational Sciences for Energy Research" (Grant no. 13CSER025), of the Netherlands Organization for Scientific Research (NWO), which was co-financed by Shell Global Solutions International B.V. G.M.C. was also financially supported by NWO. S.B. and D.Z. acknowledge financial support from ERC Starting Grant No. 335078 COLOURATOMS. T.A. acknowledges a post-doctoral grant from the Research Foundation Flanders (FWO, Belgium). 
C.B.M and Y.W. acknowledge support for materials synthesis from the Office of Naval Research Multidisciplinary University Research Initiative Award ONR N00014-18-1-2497.
\subsection*{Author contributions}
 A.v.B. initiated the investigation of icosahedral order in binary crystals under spherical confinement and supervised D.W. and E.B.v.d.W. M.D. initiated the simulation study on the binary iQC clusters in spherical confinement and supervised T.D. and G.M.C. T.D. performed computer simulations of the spontaneous nucleation of binary iQC clusters. D.W. synthesised the SPs and obtained, together with T.A., the electron tomography. D.Z. performed SSR tomographic reconstruction and T.A. performed manual segmentations, under the supervision of S.B. Y.W. synthesised NCs under the supervision of C.B.M. E.B.v.d.W., T.D. and G.M.C. developed the BOP analysis of the Laves phases, the iQC clusters and the cluster criterion to track the nucleation of the iQC clusters. D.W., T.D., E.B.v.d.W., M.D. and A.v.B. analysed the results. A.v.B., M.D., D.W., T.D. and E.B.v.d.W. co-wrote the manuscript. All authors discussed the text and interpretation of the results.

\subsection*{Additional information}

\subsection*{Competing interests}
The authors declare no competing interests.

%\bibliography{Binary_Ico_Quasicrystal}
\printbibliography

\subsection*{Extended data}
\doublespacing
\renewcommand{\figurename}{Extended Data Figure}% Renew the Fig caption 
\setcounter{figure}{0} % reset the numbering of figures %\renewcommand{\headrulewidth}{0.2pt}
\clearpage

\begin{figure}[!t]
\includegraphics[width=0.80\textwidth]{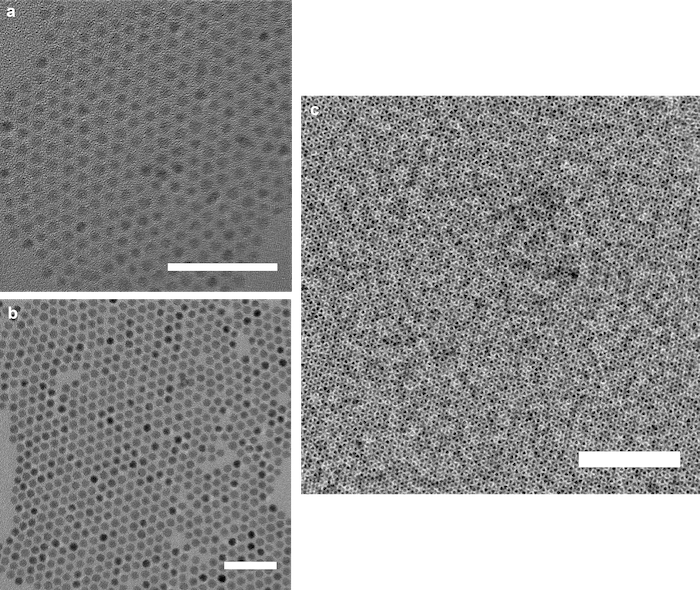}
\centering
\caption{
\textbf{TEM images of binary NCs and self-assembled binary NC superlattices.}
TEM images of 
a) 7.7 nm CdSe (a total polydispersity of 2\%; 5.2 nm core diameter), 
b) 9.9 nm PbSe NCs (a total polydispersity of 1\%; 7.6 nm core diameter) 
and c) self-assembled binary NC superlattices in bulk. Scale bars, a-b) 50 nm, c) 200 nm. 
\label{ExtendedFig_NCs_BNSL}}
\end{figure} 

\begin{figure}[!t]
\includegraphics[width=0.8\textwidth]{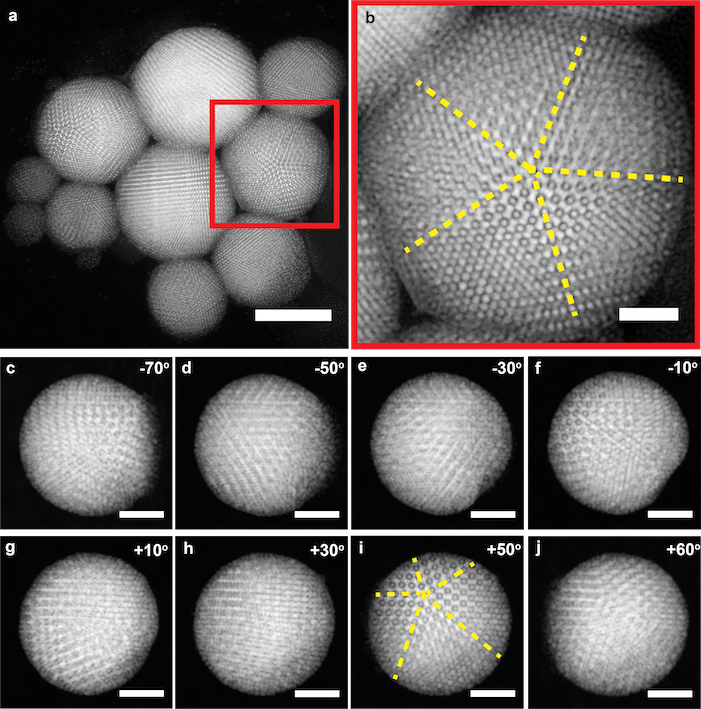}
\centering
\caption{
\textbf{HAADF-STEM images of self-assembled binary SPs showing five-fold symmetry.}
a) Overview of self-assembled binary SPs. 
b) A SP with a diameter of 293 nm showing five-fold symmetry, marked with yellow dashed lines. 
c-j) a SP with a diameter of 158 nm at different tilting angles. The five-fold symmetry can be viewed at a tilting angle of +50$^\circ$. 
Scale bars: a) 200 nm, b) 50 nm, c-j) 50 nm, respectively.
\label{ExtendedFig_ExtendedFig_Exp_SPs_overview_and_tilting_v2}}
\end{figure}

\begin{figure}[!t]
\includegraphics[width=0.8\textwidth]{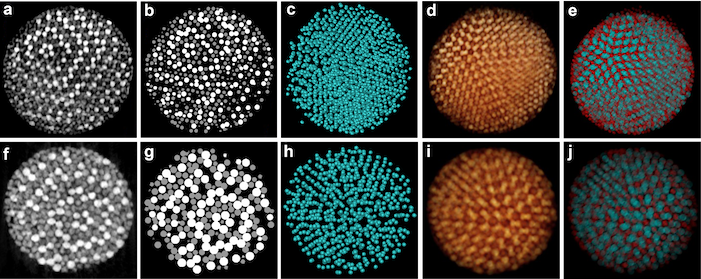}
\centering
\caption{
\textbf{Tomographic reconstruction of two self-assembled binary SPs.}
Tomographic reconstructed orthoslices through the middle part of the reconstructed volumes using a,f) SIRT and b,g) SSR. 
c,h) Manually segmented \textit{L} species (PbSe NCs). 
3D representation of the reconstructed volumes using d,i) SIRT and e,j) SSR, for two binary SPs with a diameter of a-e) 170 nm and f-j) 110 nm, respectively. 
a,b,f,g) Bright spheres and dim spheres represent \textit{L} species and \textit{S} species, respectively. e,j) \textit{L} and \textit{S} species are coloured with cyan and red, respectively. 
The transparency of the particles in d,e,i,j were increased for visual clarity.
\label{Extended_Fig_reconstruction}}
\end{figure}

\begin{figure}[!t]
\includegraphics[width=0.8\textwidth]{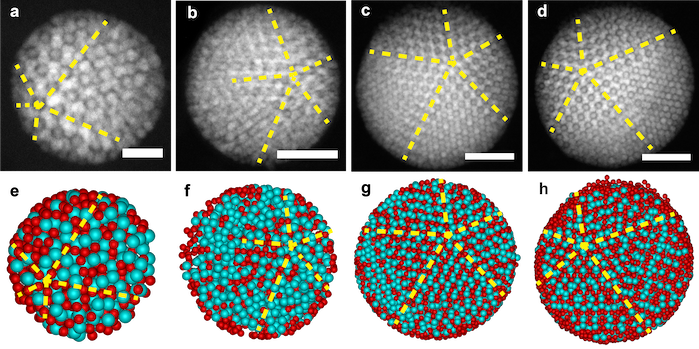}
\centering
\caption{
\textbf{Electron tomography analysis of self-assembled binary SPs of different sizes.}
a,b,c,d) 2D HAADF-STEM images and e,f,g,h) corresponding reconstructed 3D representations of self-assembled binary SPs with diameters of a) 80 nm, b) 132 nm, c) 167 nm and d) 170 nm. 
Particles in cyan and red represent PbSe and CdSe NCs, respectively. 
Five-fold symmetry was illustrated by yellow dashed lines. 
Scale bars: a) 20 nm, b-d) 50 nm, respectively. 
For the number of particles in the self-assembled SPs, see Supplementary Table 1. 
\label{ExtendedFig_TEM_five_fold_gallery}}
\end{figure}

\begin{figure}[!t]
\includegraphics[width=0.8\textwidth]{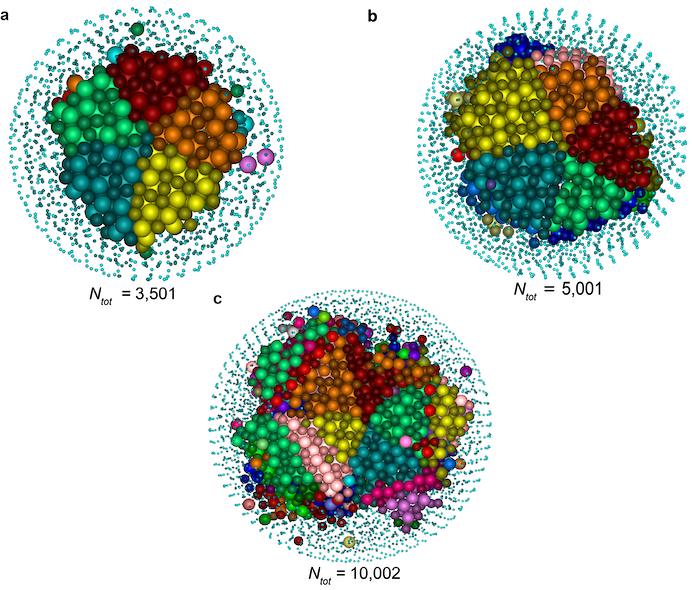}
\centering
\caption{
\textbf{SPs of different system sizes as obtained from MD simulations.}
The simulations are performed on system sizes of $N_{\small tot}$ = 3,501, 5,001 and 10,002 spheres, respectively.
%with number densities $\rho\sigma_\textrm{\scriptsize av}^{\small 3}$ ($\rho\sigma_\textrm{\scriptsize HS, eff}^{\small 3}$) 0.802 (0.966), 0.809 (0.974) and 0.831 (1.001). See Supplementary Section 5 for more details. 
Binary iQCs are observed to spontaneously crystallize for all of them. Crystalline particles belonging to the iQC cluster are identified using BOPs. The different crystalline domains are shown in different colours. For crystalline and domain classification, see Supplementary Section 4. The \textit{S} particles in each domain are coloured in a darker shade of the same colour as the \textit{L} particles. Particles not belonging to the iQC cluster are reduced in size for visual clarity.
\label{ExtendedFig_Simulated_Colored}}
\end{figure}

\begin{figure}[!t]
\includegraphics[width=0.8\textwidth]{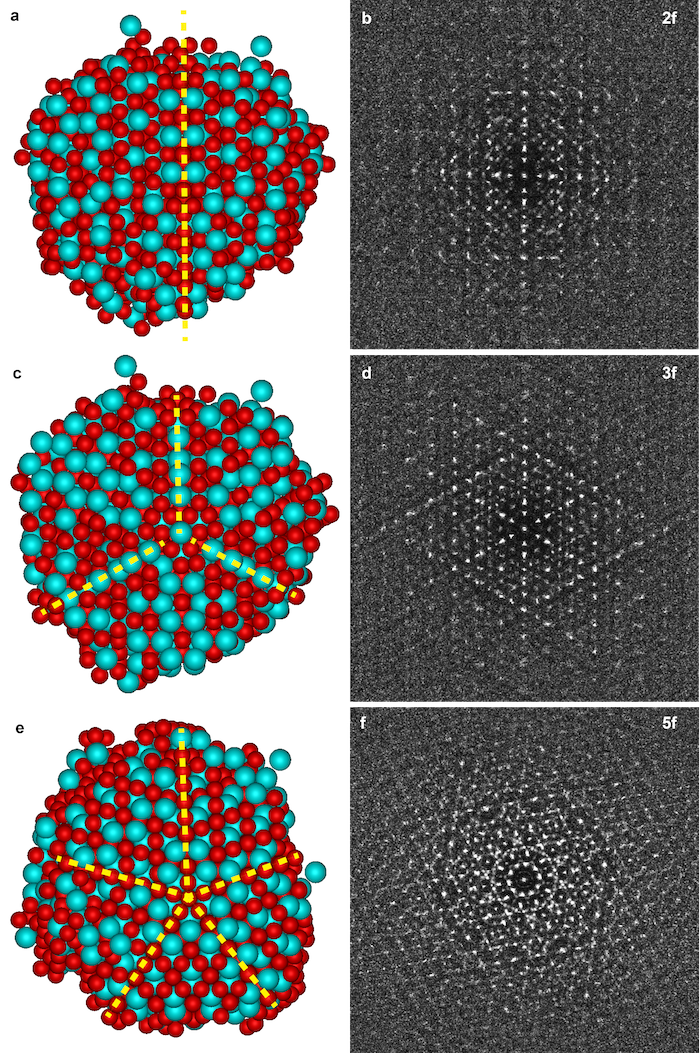}
\centering
\caption{
\textbf{Cross-section views of computer rendered representations of a simulated SP  exhibiting two-, three- and five-fold symmetries.}
a,c,e) Real-space configurations of the simulated SP composed of $N_{\small tot}$ = 5,001 spheres and b,d,f) corresponding calculated diffraction patterns of a, c and e, viewed along two- (2f), three- (3f) and five-fold (5f) axes. The diffraction pattern is calculated based on the time-averaged coordinates of the particles belonging to the iQC cluster identified using BOPs described in Supplementary Section 4.
\label{ExtendedFig_2-3-fold_symmetries_FFT}}
\end{figure} 

\begin{figure}[!t]
\includegraphics[width=0.8\textwidth]{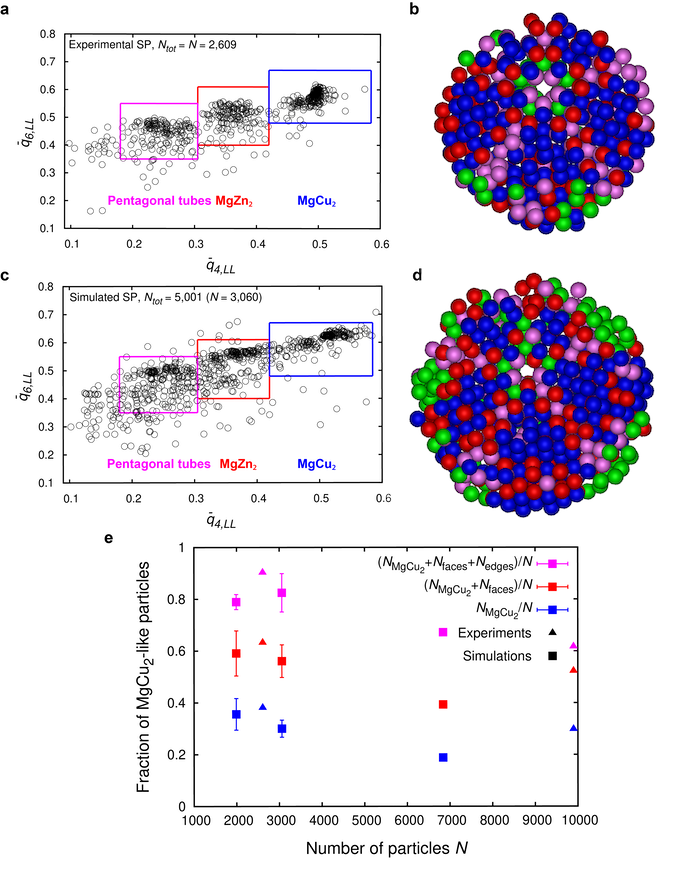}
\centering
\caption{
\textbf{Local structural symmetries of \textit{L} species in SPs.} 
Scatter plot of the average BOPs of the \textit{L} particles in the $\overline{q}_{4,LL}$ - $\overline{q}_{6,LL}$ plane of the crystal structure of a) an experimental and 
c) a simulated SP. For the simulated SP, only the BOP values of the $N$ particles in the core are plotted (see Supplementary Section 3).
The blue and red boxes signify the BOP values that we used to identify the MgCu$_2$-like and MgZn$_2$-like local symmetry, respectively, of the particles in the final structure. 
The purple box represents the symmetry of the pentagonal tubes (or icosahedral chains) that run through the SP. 
These pentagonal tubes are formed by the edges of five neighbouring tetrahedral domains. 
Cross-section views of final configurations of the corresponding b) experimental SP ($N$ = 2,609, 110 nm) and d) simulated SP ($N_{tot}$ = 5,001), in which the particles are coloured accordingly (\textit{i.e.} MgZn$_2$-like, MgCu$_2$-like particles, and pentagonal tubes are coloured in red, blue and purple, respectively). 
All other particles (not belonging to either of the three boxes) are coloured green. 
It is observed that the tetrahedral domains in the icosahedral SP are composed of particles with MgCu$_2$-like symmetry (blue). The shared faces of adjacent tetrahedral MgCu$_2$ domains show MgZn$_2$-like symmetry (red). 
e) Fraction of MgCu$_2$-like particles for different system sizes: experimental with $N$ = 2,609 (110 nm) and 9,898 particles (187 nm)
and simulated SPs with $N$ = 1,990, 3,060 and 6,870 ($N_{tot}$ = 3,501, 5,001 and 10,002) particles. 
The different estimates of the MgCu$_2$-like fraction are based on the different grouping of particle symmetries as described in the legend. The error bars are obtained from three independent simulations. 
\label{ExtendedFig_BOP}}
\end{figure}

\begin{figure}[!t]
\includegraphics[width=0.8\textwidth]{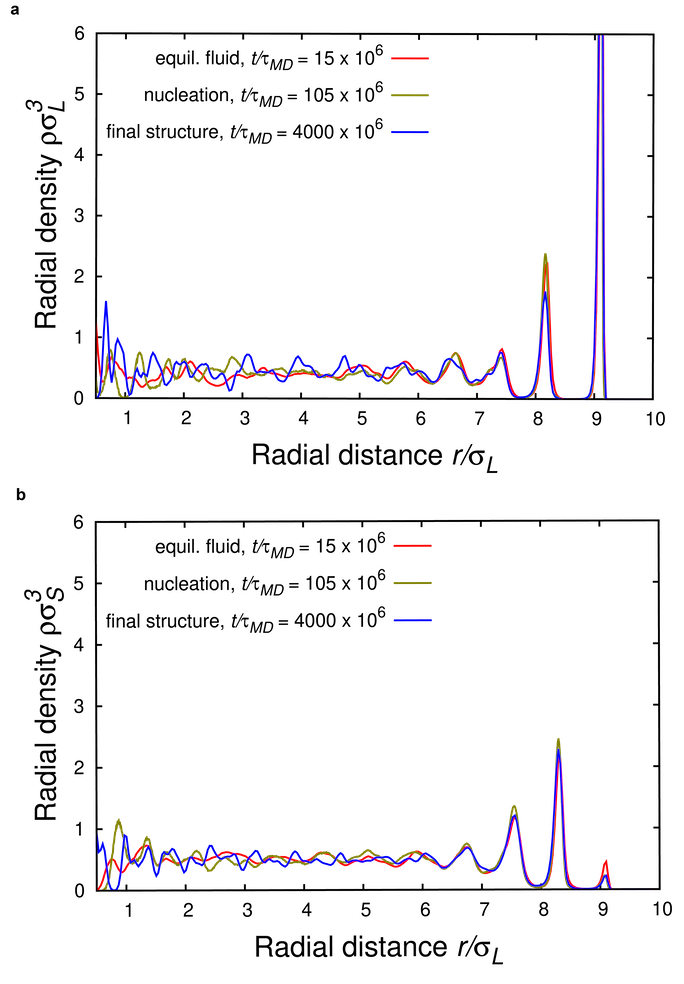}
\centering
\caption{
\textbf{Species density profiles of the highly structured fluid phase.}
Radial density profiles of a) $L$ species and b) $S$ species of a binary mixture of $N_{tot}$ = 5,001 HS-like simulated particles in spherical confinement in the equilibrated fluid phase (Fig. \ref{Figure_5_nucleation}b), at the onset of nucleation (Fig. \ref{Figure_5_nucleation}d), and the final SP structure (Fig. \ref{Figure_5_nucleation}g). Strong layering in the outermost shells is observed early in the simulation. The radial density profiles are averaged over 1,000 configurations.
\label{ExtendedFig_layering_and_correlations}}
\end{figure} 

%\begin{figure}[!t]
%\includegraphics[width=0.8\textwidth]{figures/ExtendedFig_Radial_Composition_Profile}
%\centering
%\caption{
%\textbf{Radial composition profile.} 
%Radial composition (number fraction of small species) profile of a binary mixture of $N_{\small tot}$ = 5,001 HS-like simulated particles in the (i) equilibrated fluid phase (Fig. \ref{Figure_5_nucleation}b), (ii) at the onset of nucleation (Fig. \ref{Figure_5_nucleation}d), and (iii) the final SP structure (Fig. \ref{Figure_5_nucleation}g). The radial composition profiles are averaged over 1000 configurations.
%\label{ExtendedFig_Radial_Composition_Profile}}
%\end{figure} 

\end{document}

%\linenumbers
%\maketitle

%%%%%%%%%%%%%%%%%%%%%%%%%%%%%%%%%%%%%%%%%%%%%%%%%%%%%%%%%%%%%%%
\section*{Supplementary Methods}

\section{Chemicals}
Chemicals used were: dextran from Leuconostoc mesenteroides (Sigma Aldrich, 
mol.\ wt.\ 1,500,000-2,800,000), cyclohexane (Sigma Aldrich, $\ge$99.8\%), \textit{n}-hexane (Sigma
Aldrich, $\ge$99.5\%), 1-octadecene (1-ODE, Sigma Aldrich, 90\%) Sodium dodecyl
sulfate (SDS, Sigma Aldrich, $\ge$99.0\%), oleic acid (OA, Sigma Aldrich, $\ge$99.0\%),
isopropanol (Sigma Aldrich, $\ge$99.5\%), diphenylphosphine (Sigma Aldrich, $>$90\%), trioctylphosphine (TOP,
Sigma Aldrich, 97\%), selenium shot (Se, Alfa Aesar, 99.999\%), lead (II) oxide (PbO, Sigma Aldrich, 99.999\%), cadmium oxide (CdO, Sigma Aldrich, $\ge$99.99\%), toluene (Sigma Aldrich, anhydrous, 99.8\%), diethylene glycol (DEG, Sigma Aldrich, 99\%).
For de-ionized water (DI H$_2$O) a
Millipore Direct-Q UV3 reverse osmosis filter apparatus was used (\SI{18}{M \ohm} at
\SI{25}{\degree C}).

\section{Nanocrystal (NC) syntheses and self-assembled supraparticles (SPs)} 
\subsection{CdSe NCs}
7.7 nm CdSe NCs including ligands (a total polydispersity of 2\% by counting 130 CdSe NCs in TEM; 5.2 nm core diameter; Extended Data Fig. 1a) were synthesised by cation exchange from 5.2 nm PbSe template based on reported recipes \autocite{Pietryga2008}. 750 mg of CdO, 5.0 mL of OA and 25 mL 1-ODE were mixed and degassed at \SI{105}{\degree C} for 30 minutes. Then the mixture was heated to \SI{250}{\degree C} in a nitrogen atmosphere to get a clear solution. The solution was kept at \SI{250}{\degree C} for 20 minutes, then allowed to cool down to \SI{105}{\degree C}. It was degassed again for 1 hour and then heated to \SI{210}{\degree C}. 150 mg of PbSe NCs were dispersed in 3.0 mL of toluene and swiftly injected into the reaction flask. The solution turns red in a few seconds, indicating the formation of CdSe NCs. The reaction was kept at \SI{210}{\degree C} for 45 minutes and then cooled down to room temperature (RT). The as-synthesised CdSe NCs were purified with isopropanol three times, dispersed in \textit{n}-hexane and stored in a glove box. 

\subsection{PbSe NCs}
9.9 nm PbSe NCs including ligands (a total polydispersity of 1\% by counting 170 PbSe NCs in TEM; 7.6 nm core diameter; Extended Data Fig. 1b) were synthesised following reported recipes \autocite{steckel2006}. Specifically, 0.895 g of PbO, 3.0 mL of OA, and 20 mL of 1-ODE were mixed and degassed at \SI{105}{\degree C} for 1.5 hours. Then the solution was heated to \SI{105}{\degree C} in a nitrogen atmosphere. \SI{71} {\micro L} of diphenyl phosphine was mixed with 8.0 mL of trioctylphosphine-selenium (TOP-Se) solution (1 M). The mixture was swiftly injected into the reaction flask. The NCs were allowed to grow at \SI{105}{\degree C} for 10 minutes and then quenched by an ice bath. The as-synthesised PbSe NCs were purified with isopropanol three times, dispersed in \textit{n}-hexane and stored in a glove box.

\subsection{Experimental self-assembled SPs}
\begin{table}[htp]
\centering
\begin{tabular}{c c c c } \hline
        \hline
$D$ (nm)&$N$&$N_L$&$N_S$\\[0.1 cm]
		\hline	
80&832&300&532\\
\hline
110&2,609&961&1,648\\
\hline
132&2,631&1,323&1,308\\
\hline  
167&7,767&3,037&4,730\\
\hline 
170&9,978&2,809&7,169\\
\hline 
187&9,898&3,267&6,631\\
\hline 
\end{tabular}
\caption{Details of experimental SPs discussed in this work, from a binary mixture of NCs with a size ratio of 0.78 under spherical confinement. $D$, $N$, $N_L$ and $N_S$ are the diameter, number of binary species, $L$ species (PbSe NCs) and $S$ species (CdSe NCs), respectively.
\label{exp_SP_number}}
\end{table}

%\section{Electron tomography.}
%All Electron tomography experiments were performed on ......
%A Fischione 2020 single tilt
%For SP with a diameter of 80 nm, 110 nm, 132 nm, 167 nm, 170 nm and 187 nm, tilt series were performed within a tilt range:  
%from -\SI{72}{\degree} to +\SI{74}{\degree}, 
%from -\SI{76}{\degree} to +\SI{70}{\degree}, 
%from -\SI{78}{\degree} to +\SI{68}{\degree}, 
%from -\SI{72}{\degree} to +\SI{74}{\degree}, 
%from -\SI{70}{\degree} to +\SI{68}{\degree} and 
%from -\SI{70}{\degree} to +\SI{74}{\degree}, respectively. 
%Increment in all experiments were set to be \SI{2}{\degree}. 

%For SP with a diameter of 80 nm,
%For SP with a diameter of 132 nm,
%For SPs with a diameter of 187 nm,

%\section{Manual segmentation}
%\hl{In order to check the accuracy of SSR tomographic reconstruction, manual segmentation of \textit{L} particles (\textit{i.e.} PbSe NCs) was carried out for experimental SPs with a diameter of 170 nm and 110 nm, respectively. To be added by T.A.}
%Figure will be added (orthoslice, SSR and manual segmented in one fig).

\section{\label{bop_ll}Bond-orientational order parameter analysis of the local symmetry of the particles} 
%We use the average BOP \autocite{lechner2008accurate} to characterise the correlation between only the \emph{L} species in our SPs (Supplementary Fig. \ref{SI_Fig_Bond_Order_Laves_Phases}). In particular, we use the $\overline{q}_{\small 4}$ and $\overline{q}_{\small 6}$ order parameters to identify the crystal structure of our SPs. Here, $\overline{q}_{\small l}$ of a particle $i$ is defined as $\overline{q}_{\small l} (i) = \sqrt{\frac{4\pi}{2l+1}\sum_{m=-l}^{l} \vert \overline{q}_{l,m}(i)\vert^2}$ where $\overline{q}_{l,m}(i) = \frac{1}{N_{b}(i)+1}\sum_{k=0}^{N_{b}(i)} q_{l,m}(k)$ with $N_{b}(i)$ denoting the number of (large) neighbours of \hl{(\textit{L})} particle $i$ and $k = 0$ represents the particle $i$ itself. Here, $q_{l,m}(i)$ is the traditionally defined normalised 3D bond order parameter for particle $i$ \autocite{steinhardt1983bond}. The distinct OP clouds on the $\overline{q}_{\small 4}$ -- $\overline{q}_{\small 6}$ plane assist us in identifying the local environment of the NCs in our SPs. This analysis is performed for both our simulated and experimental SPs.

\noindent The local symmetry of the $L$ particles in the SPs was characterised using averaged bond-orientational order parameters (BOPs)~\autocite{lechner2008accurate}. In particular, the $\overline{q}_{4}$ and $\overline{q}_{6}$ order parameters were used. $\overline{q}_{l}$ of a particle $i$ is defined as:
\begin{linenomath*}
\begin{equation*}
 \overline{q}_{l} (i) = \sqrt{\frac{4\pi}{2l+1}\sum_{m=-l}^{l} \vert \overline{q}_{l,m}(i)\vert^2},
 \end{equation*}
 \end{linenomath*}
 where: 
 \begin{linenomath*}
 \begin{equation*}
 \overline{q}_{l,m}(i) = \frac{1}{N_{b}(i)+1}\sum_{k=0}^{N_{b}(i)} q_{l,m}(k),
 \end{equation*}
 \end{linenomath*}
with $N_{b}(i)$ denoting the number of ($L$) neighbours of ($L$) particle $i$ and $k = 0$ represents the particle $i$ itself. $q_{l,m}(i)$ is the traditionally defined normalised 3D bond order parameter for ($L$) particle $i$~\autocite{steinhardt1983bond}. 
Nearest neighbours were determined by locating all particles within a cut-off distance $r_c$ from particle $i$, where $r_c$ corresponded to the first minimum of the radial distribution function $g_{LL}(r)$. As particles close to the SP surface have a lower number of nearest neighbours and therefore have a lower BOP value, particles closer than $2\sigma_L$ to the spherical confinement in simulations were not presented in the scatter plot and ignored in the estimation of the threshold BOP values which were used to characterise the local structural environment. For the experimental data sets, the center of mass of the $L$ particles was defined, after which a sphere from this center was defined with a radius such that a minimal of two layers of particles were excluded. The fraction of MgCu$_2$-like particles is calculated without accounting for these `surface' particles.

As a reference, the scatter plots of the $\overline{q}_{4,LL}$ and $\overline{q}_{6,LL}$ parameters of the $L$ particles in the three Laves phases were calculated for $NVT$ equilibrated Laves phases at coexistence density $\rho\sigma_\textrm{\scriptsize av}^3$ = 0.953 (Supplementary Fig. \ref{SI_Fig_Bond_Order_Laves_Phases}). By subsequent colouring of the particles in relation to their position in the $\overline{q}_{6,LL}$--$\overline{q}_{4,LL}$ scatter plot, their local symmetry was identified. To distinguish the local environment corresponding to the MgZn$_2$-like and MgCu$_2$-like symmetries, the following regions in the scatter plot were defined as $ 0.305 < \overline{q}_{4,LL} < 0.42 $ and $ 0.4 < \overline{q}_{6,LL} < 0.61 $ (MgZn$_2$-like), and $ 0.42 < \overline{q}_{4,LL} < 0.585 $ and $ 0.48 < \overline{q}_{6,LL} < 0.67 $ (MgCu$_2$-like). In addition, we used an additional region to colour the $L$ particles that are part of the pentagonal tubes: $ 0.18 < \overline{q}_{4,LL} < 0.305 $ and $ 0.35 < \overline{q}_{6,LL} < 0.55$.

\section{\label{bop_nucl}Tracking the nucleation and growth of the iQC cluster in spherical confinement} 
\noindent We used a different approach to detect the nucleation of crystalline clusters in the supersaturated fluid in our simulation trajectories. Here we used a dot product 
$d_{l,\alpha\beta}(i,j) =\sum\limits_{m=-l}^l q^{\alpha}_{l,m}(i) q^{*\beta}_{l,m}(j)$ correlation between particle $i$ and its neighbour $j$ for all three species combinations, with $\alpha,\beta$ denoting the type of species, to identify particles belonging to the iQC crystalline cluster. Nearest neighbours $j$ were determined by locating all particles within a cut-off distance $r_c$ from particle $i$, where $r_c$ corresponds to the first minimum of the radial distribution function $g_{\alpha\beta}(r)$. For the (i) \textit{L}-\textit{L} species correlation, we used $d_{6,LL} >  0.4$, (ii) \textit{S}-\textit{S} species correlation, $-0.23 < d_{6,SS} < -0.04$ and (iii) \textit{S}-\textit{L} species correlation  $-0.46 < d_\textrm{6,LS} < -0.1$. If the dot product $d_{l,\alpha\beta}(i,j)$ between the BOPs of two particles is in the range of the cut-off values specified (based on the species types), they share a ``solid-like'' bond. If the total number of solid-like bonds of a particle $i$ with its neighbours was greater than a threshold value $\xi_c$, then particle $i$ was crystalline. For a \textit{L} particle, $\xi_c^L$ = 10 and for a \textit{S} particle, $\xi_c^S$ = 9. The above criteria were sufficient to distinguish the crystalline particles from the surrounding fluid.  Additionally, to distinguish the tetrahedral domains from each other, we employed a dot product $\overline{d}_{6,\alpha\beta}(i,j)$ correlation between two neighbouring \textit{L} crystalline particles $i$ and $j$, and identified $i$ and $j$ as belonging to the same crystalline domain if $\overline{d}_{6,LL}(i,j) > 0.8$. Here $\overline{d}_{l,\alpha\beta}(i,j)$ has the same form as defined above, but is calculated with the average BOPs \autocite{lechner2008accurate}. $\overline{q}_{l,m}(i)$ is defined similarly as in Supplementary~\ref{bop_ll}. The \textit{S} particles were assigned domain IDs corresponding to their closest crystalline \textit{L} particle neighbour.

\section{Effective diameter of an equivalent hard-sphere system from WCA theory}
\noindent Historically, the soft WCA potential has been employed as a perturbation to a reference hard sphere (HS) system, and can be mapped to an equivalent HS system through the determination of an \emph{effective diameter}. This effective diameter may be calculated through a Taylor expansion of the reduced excess free-energy density of the WCA system $f = \beta F^{ex}/V$, where $\beta = 1/k_B T$, in powers of the difference in the Boltzmann factor of the WCA system and a hard sphere system $\Delta e({\bf r})$:
\begin{linenomath*}
\begin{equation*}
f = f_d + \int \frac{\partial f}{\partial e({\bf r})}\bigg\vert_{e = e_d} \Delta e({\bf r}) \textrm{d}{\bf r} + \frac{1}{2} \int \int \frac{\partial^2 f}{\partial e({\bf r}) \partial e({\bf r}^{\bf '})}\bigg\vert_{e = e_d} \Delta e({\bf r}) \Delta e({\bf r}^{\bf '}) \textrm{d}{\bf r} \textrm{d}{\bf r}^{\bf '} + ...
\label{exp_fe}
\end{equation*}
\end{linenomath*}
Here, $\Delta e({\bf r}) = e_0({\bf r}) - e_d({\bf r})$ where $e_0({\bf r}) = \exp\left[-\beta \phi (r)\right]$ is the Boltzmann factor of the WCA system with $\phi(r)$ the WCA potential and $e_d({\bf r})$ the Boltzmann factor of the HS system. A detailed discussion of the effective diameter $\sigma^{\footnotesize eff}$ estimation is found in Ref.~\cite{hansen1990theory}, where in essence, equating the free-energy density of the WCA and HS systems (and ignoring the high order terms) results in the effective diameter $\sigma^{\footnotesize eff}$ of the WCA system being defined as follows:
\begin{linenomath*}
\begin{equation*}
\sigma^{\footnotesize eff} = \int_0^\infty \left(1 - \exp\left[-\beta \phi(r)\right]\right) \textrm{d}r .
\label{eff_dia}
\end{equation*}
\end{linenomath*}
For the WCA system $T^* = k_BT/\epsilon = $ 0.2, the effective diameter $\sigma_{\alpha}^{\footnotesize eff} = $  1.0639 $\sigma_{\alpha}$, where $\sigma$ is the WCA diameter as defined in Methods and $\alpha$ denotes the type of species (\emph{i.e.} $L$,$S$). Our nucleation simulations were performed in a constant spherical volume comprising of a binary mixture of WCA spheres with a size ratio of 0.78, in order to speed up equilibration. The number densities and compositions which yield the simulated SPs discussed in this work, and the corresponding effective packing fractions of an equivalent binary HS mixture with a size ratio of 0.78, are listed in Supplementary Table~\ref{reproduce}.

\begin{table}[htp]
\centering
\begin{tabular}{c c c c c c c} \hline
        \hline
$N_{tot}$&$N_L$&$N_S$&$R_{SC}$&$\rho\sigma_{av}^3$&$\rho\sigma_{av}^{3^{(c)}}$&$\eta_\textrm{\tiny HS,eff}$\\[0.1 cm]
		\hline	
3,501&1,366&2,135&8.913&0.802&0.957&0.506\\
\hline
5,001&2,001&3,000&10.036&0.809&0.945&0.510\\
\hline
1,0002&4,001&6,001&12.532&0.831&0.941&0.524\\
\hline  
\end{tabular}
\caption{Simulation details for nucleating the simulated SPs discussed in this work, from a binary mixture of WCA particles with a size ratio $\sigma_S/\sigma_L$ = 0.78 under spherical confinement. $N_{tot}$, $N_L$ and $N_S$ are the total number of particles, number of large WCA spheres and number of small WCA spheres respectively. $R_{SC}$ denotes the radius of the spherical shell, $\rho\sigma_{av}^3$ = $\frac{3N_{tot}\sigma_{L}^3}{4\pi R_{\small SC}^3}\left(\frac{N_L}{N_{tot}}+q^3\frac{N_S}{N_{tot}}\right)$ is the reduced number density of the spherical volume, $\rho\sigma_{av}^{3^{(c)}}$ is the number density of the compressed SP, and $\eta_\textrm{\tiny HS,eff}$ = $\frac{\pi}{6}\rho\sigma_{av}^3 \times \frac{\sigma_{L}^{\footnotesize eff^3}}{\sigma_{L}^3}$ is the effective packing fraction of an equivalent binary HS mixture with a diameter ratio of 0.78.
\label{reproduce}}
\end{table}

\clearpage
\newpage

\section*{Supplementary Figures}

\newpage 

\begin{figure}[!t]
\includegraphics[width=0.8\textwidth]{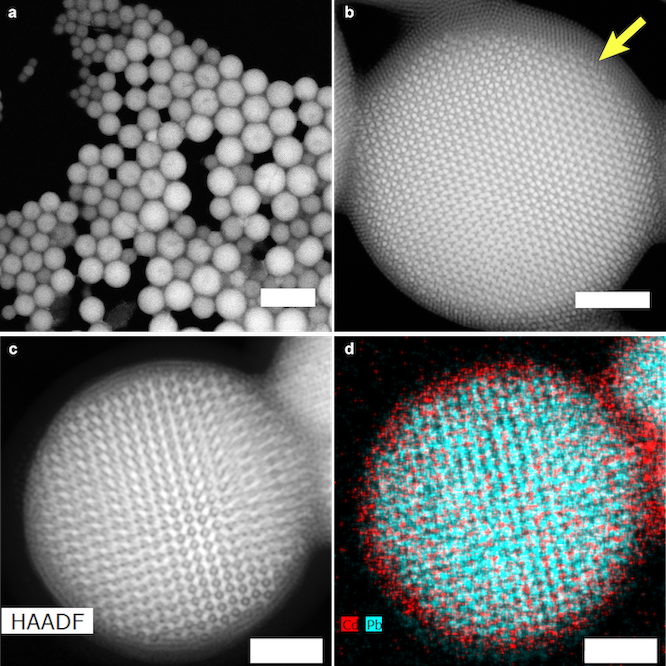}
\centering
%\internallinenumbers
\caption{
\textbf{Influence of stochiometry of binary NCs on self-assembled supraparticles (SPs).} 
a) Overview and b) a zoomed-in view of self-assembled binary SPs without icosahedral geometry, showing a MgZn$_2$ crystal structure (viewed along the [0001] projection) in coexistence with shells containing CdSe NCs.
c) High angle annular dark field-scanning transmission electron microscopy (HAADF-STEM) image and d) elemental distribution superimposition of Cd (red) and Pb (cyan) in a SP with a diameter of 246 nm, viewed along the [11$\bar{2}$0] projection of the MgZn$_2$ crystal structure. 
Pb was found in the core only, while Cd was present in both the core and shell of the SP. 
Note that the experimental parameters are the same as those used to obtain icosahedral quasicrystals (iQCs) except that an excess amount (10\% by weight) of CdSe NCs was used. 
Curved outer layers composed of close-packed CdSe NCs can be viewed clearly in the self-assembled SPs, which are indicated for one SP by a yellow arrow (b). 
Scale bars, a): 200 nm, b): 100 nm, c-d): 60 nm.
\label{SI_Fig_off_stochiometry}}
\end{figure} 
%\clearpage
\begin{figure}[!t]
\includegraphics[width=1\textwidth]{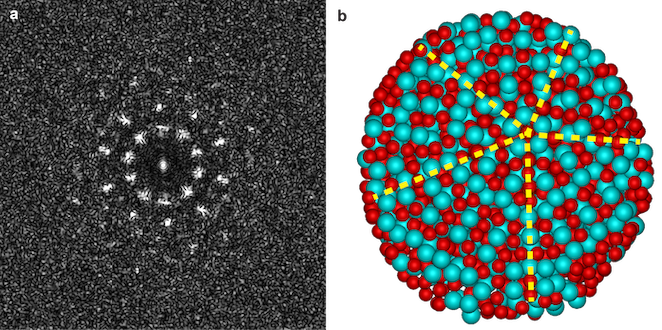}
\centering
\caption{
\textbf{Calculated diffraction pattern of an experimental binary icosahedral SP.} 
a) Calculated diffraction pattern and b) real-space coordinates of an experimental binary SP containing 2,609 NCs, viewed along a five-fold (5f) axis. For an interactive view, see Supplementary Data 4.
\label{SI_Fig_exptl_fft}}
\end{figure}

\begin{figure}[!t]
\includegraphics[width=1\textwidth]{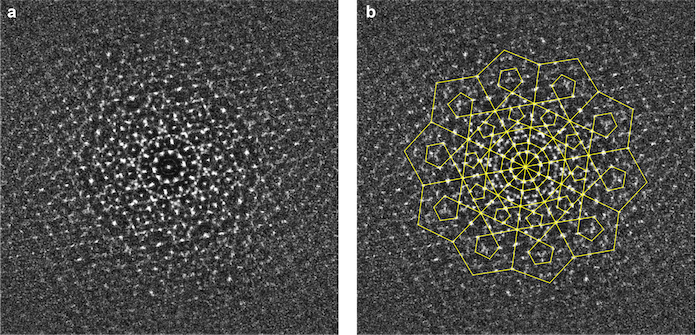}
\centering
\caption{
\textbf{Calculated diffraction pattern of a simulated binary icosahedral SP.} 
a) The calculated diffraction pattern of a simulated binary SP containing $N_{\small tot}$ = 5,001 ($N$ = 3,060) particles, showing five-fold orientational symmetry. The diffraction is calculated from time-averaged coordinates of the particles belonging to the iQC cluster identified using bond-orientational order parameters (BOPs) described in Supplementary Section 4. The particle coordinates are time-averaged over 10,000 configurations.
b) The same pattern with an overlay pointing out the presence of pentagonal symmetry in the pattern, corresponding to an icosahedral quasicrystalline structure. The yellow lines highlight the presence of inflating pentagons in the pattern, which is a direct consequence of the non-crystallographic five-fold orientational symmetry, depicting quasiperiodicity.
\label{SI_Fig_full_patterns}}
\end{figure} 
%\clearpage

\begin{figure}[!t]
\includegraphics[width=1\textwidth]{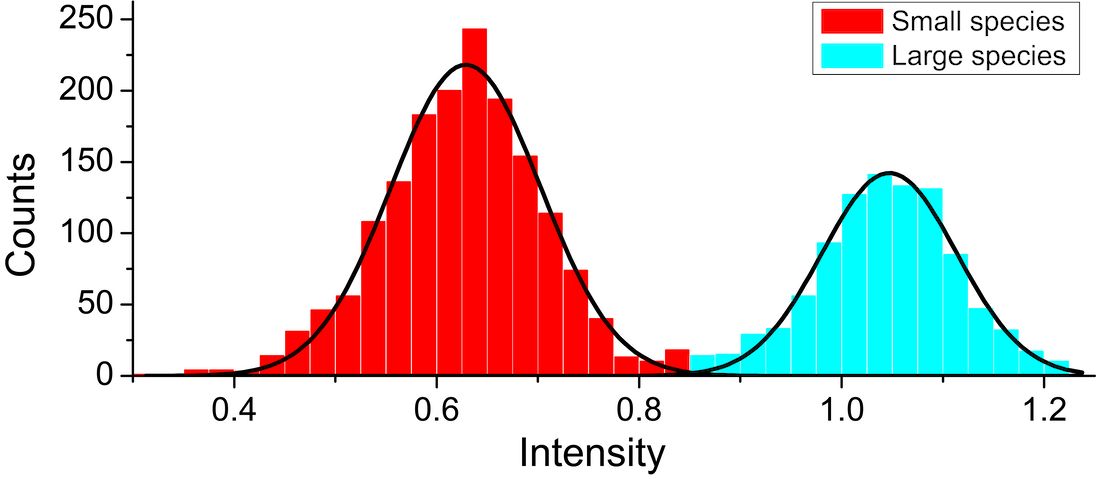}
\centering
\caption{
\textbf{Pixel intensity distributions of reconstructed particles in a binary SP.} 
Distribution of the summed pixel intensities of the particles in the 110 nm binary SP. 
The black curves are two Gaussian distributions fitted to the intensity distributions, from which threshold values were determined to differentiate between the CdSe (red) and PbSe (cyan) NCs. The overlap between the two Gaussian curves gives an estimate for the percentage of particles that might be misidentified ($<1\%$). 
\label{SI_Fig_tomo2_int_reassigning}}
\end{figure} 
%\clearpage

\begin{figure}[!t]
\includegraphics[width=1\textwidth]{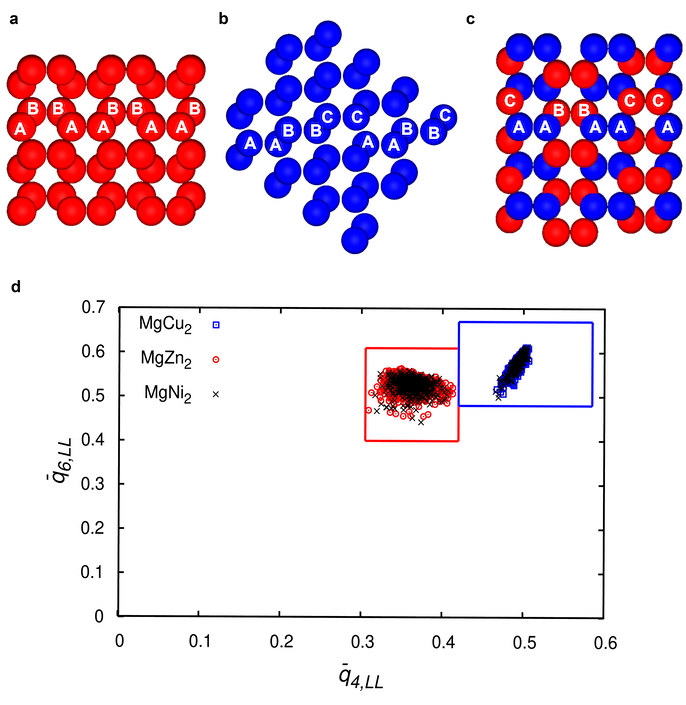}
\centering
\caption{
\textbf{Bond orientational order parameters (BOP) of the three Laves phases.} 
Computer renderings of the $L$ particles in the perfect crystal structures of the three Laves phases:
a) MgZn$_2$, 
b) MgCu$_2$ and 
c) MgNi$_2$. 
The stacking of the \textit{L} particle-dimers in the MgZn$_2$, MgCu$_2$ and MgNi$_2$ phases is ...\textit{AABB}..., ...\textit{AABBCC}... and ...\textit{AABBAACC}..., respectively.
d) Scatter plot of the BOP values of the \textit{L} particles in the $\overline{q}_{6}$ - $\overline{q}_{4}$ plane of the three Laves phases MgCu$_2$ (blue squares), MgZn$_2$ (red circles) and MgNi$_2$ (black crosses), where the structures were equilibrated at melting density $\rho\sigma_\textrm{\scriptsize av}^{\small 3}$ = 0.953. The effective binary HS packing fraction $\eta_\textrm{\tiny HS, eff}$ = 0.601. In order to assign local symmetries in the BOP analysis of our SP structures, two regions were specified from these simulations: MgZn$_2$-like as denoted by the red rectangle and MgCu$_2$ as denoted by the blue rectangle. The $L$ particles in a-c) are coloured according to their local symmetry, which can be used to distinguish all three Laves phases.
\label{SI_Fig_Bond_Order_Laves_Phases}}
\end{figure} 
\clearpage

\printbibliography[title={Supplementary References}] %  leave out [....]: will show as References